\documentclass[useAMS,usenatbib]{mn2e}

\usepackage{graphicx}                                                           
\usepackage{amssymb}
\usepackage{epsfig}
\usepackage{epstopdf}
\usepackage{verbatim}
\usepackage{subfigure}
\usepackage{lscape}
\usepackage{longtable}
\usepackage{deluxetable}
\usepackage{rotating}
\usepackage{lscape}
\usepackage{color}
\definecolor{dkcyan}{rgb}{0.000,0.600,0.600}
\definecolor{navy}{RGB}{0,0,102}
\definecolor{royal}{RGB}{0,0,204}
\usepackage[T1]{fontenc}
\usepackage{aecompl}
\usepackage[totalwidth=17.8cm, totalheight=24.0cm]{geometry}

\newcommand{\apj}{ApJ}
\newcommand{\apjl}{ApJ}

\newcommand{\aj}{AJ}
\newcommand{\mnras}{MNRAS}
\newcommand{\aaps}{A\&AS}
\newcommand{\aap}{A\&A}
\newcommand{\pasp}{PASP}

\newcommand{\nat}{Nature}

\def\arcsec{$^{\prime\prime}$}

\title[NGC\,5813: counter-rotating nature of KDCs]{Unveiling the counter-rotating nature of the kinematically distinct core in NGC\,5813 with MUSE}
\author[Davor Krajnovi\'c et al.]{Davor Krajnovi\'c $^{1}$\thanks{E-mail:dkrajnovic@aip.de}, Peter M. Weilbacher$^{1}$, Tanya Urrutia$^{1}$, Eric Emsellem$^{2,3}$, \newauthor C. Marcella Carollo$^{4}$, Maryam Shirazi$^{4}$, Roland Bacon$^{3}$, Thierry Contini$^{5,6}$, \newauthor Beno\^it Epinat$^{5,7}$, Sebastian Kamann$^{8}$,  Thomas Martinsson$^{9}$, Matthias Steinmetz$^{1}$\\
$^{1}$Leibniz-Institut f\"ur Astrophysik Potsdam (AIP), An der Sternwarte 16, D-14482 Potsdam, Germany\\
$^{2}$ESO, European Southern Observatory, Karl-Schwarzschild Str. 2, 85748 Garching bei Muenchen, Germany\\
$^{3}$CRAL, Observatoire de Lyon, CNRS, Universit\'e Lyon 1, 9 Avenue Ch. Andr\'e, F-69561 Saint Genis Laval Cedex, France\\
$^{4}$ETH Zurich, Institute of Astronomy, Wolfgang-Pauli-Str. 27, CH-8093 Zurich, Switzerland\\
$^{5}$IRAP, Institut de Recherche en Astrophysique et Plan\'etologie, CNRS, 14, avenue Edouard Belin, F-31400 Toulouse, France\\
$^{6}$Universit\'e de Toulouse, UPS-OMP, Toulouse, France\\
$^{7}$Aix Marseille Universit\'e, CNRS, LAM (Laboratoire d'Astrophysique de Marseille) UMR 7326, 13388, Marseille, France\\
$^{8}$IAG, Institut f\"ur Astrophysik, Universit\"at G\"ottingen, Friedrich-Hund-Platz 1, D-37077 G\"ottingen, Germany\\
$^{9}$Leiden Observatory, Leiden University, P.O. Box 9513, 2300 RA Leiden, The Netherlands\\
}

\pagerange{\pageref{firstpage}--\pageref{lastpage}} \pubyear{2014}

\begin{document}
\date{Accepted 2015 April 28.  Received 2015 April 17; in original form 2015 March 11}

\maketitle

\label{firstpage}

\begin{abstract}
MUSE observations of NGC\,5813 reveal a complex structure in the velocity dispersion map, previously hinted by SAURON observations. The structure is reminiscent of velocity dispersion maps of galaxies comprising two counter-rotating discs, and may explain the existence of the kinematically distinct core (KDC). Further evidence for two counter-rotating components comes from the analysis of the higher moments of the stellar line-of-sight velocity distributions and fitting MUSE spectra with two separate Gaussian line-of-sight velocity distributions. The emission-line kinematics show evidence of being linked to the present cooling flows and the buoyant cavities seen in X-rays. We detect ionised gas in a nuclear disc-like structure, oriented like the KDC, which is, however, not directly related to the KDC. We build an axisymmetric Schwarzschild dynamical model, which shows that the MUSE kinematics can be reproduced well with two counter-rotating orbit families, characterised by relatively low angular momentum components, but clearly separated in integral phase space and with radially varying contributions. The model indicates that the counter-rotating components in NGC\,5813 are not thin discs, but dynamically hot structures. Our findings give further evidence that KDCs in massive galaxies should not necessarily be considered as structurally or dynamically decoupled regions, but as the outcomes of the mixing of different orbital families, where the balance in the distribution of mass of the orbital families is crucial. We discuss the formation of the KDC in NGC\,5813 within the framework of gas accretion, binary mergers and formation of turbulent thick discs from cold streams at high redshift. 
\end{abstract}


%
%

\section{Introduction}
\label{s:intor}

The vast majority of early-type galaxies are kinematically simple objects. Their stellar motions can easily be predicted from their light distributions, confirming they are largely oblate anisotropic systems \citep{2013MNRAS.432.1709C}. The SAURON \citep{2002MNRAS.329..513D} and ATLAS$^{\rm3D}$ \citep{2011MNRAS.413..813C} surveys showed that within their half-light radius the majority of early-type galaxies ($\sim85$ per cent ) have simple disc-like velocity maps aligned with their light distributions \citep{2008MNRAS.390...93K,2011MNRAS.414.2923K}, and that they have relatively high specific angular momentum, and, therefore, are referred to as fast rotators \citep[][]{2011MNRAS.414..888E}.

The topic of this paper is about an exception to the rule. The minority of early-type galaxies, those with irregular velocity maps \citep{2011MNRAS.414.2923K} and low specific angular momenta, referred to as slow rotators \citep{2007MNRAS.379..401E,2011MNRAS.414..888E}, can also exhibit spectacular kinematic features \citep{2004MNRAS.352..721E,2011MNRAS.414.2923K}. The most eye-catching features on velocity maps are certainly kinematically distinct cores\footnote{Sometimes they are called kinematically peculiar cores, or kinematically decoupled components or cores, but the decoupling nature of these structures is not confirmed.} (KDC), which come in a variety of forms. Often they are found in objects which have no net rotation at larger radii and only the KDC part of the galaxy rotates. In other cases, when the outer body shows rotation, the axis of rotation of the KDC and the outer body are misaligned by approximately 90\degr or 180\degr (counter-rotation). Other misalignments are also observed, but are rare \citep{1991ApJ...383..112F, 2011MNRAS.414.2923K}.  

The first object with a KDC to be discovered was NGC\,5813 \citep{1980MNRAS.193..931E, 1982MNRAS.201..975E}, the subject of this study. As kinematic studies of elliptical galaxies became more numerous in the late 1980s and early 1990s, the number of known KDCs also increased \citep{1988ApJ...327L..55F, 1988ApJ...330L..87J, 1988A&A...202L...5B, 1989ApJ...344..613F, 1994MNRAS.270..523C}. From the earlier studies it was clear that KDCs are frequent in massive ellipticals \citep{1988A&A...202L...5B}, but the ATLAS$^{\rm3D}$ survey showed that, while the KDC host galaxies span a range of masses, they make only a small fraction (7 per cent) of the total population of early-type galaxies \citep{2011MNRAS.414.2923K}. KDCs are, however, numerous among slow rotators ($\sim42$ per cent), but very rare among fast rotators \citep[$\sim2$ per cent;][]{2011MNRAS.414..888E}. 

Detection of KDCs is obviously a question of spatial resolution. \citet{2006MNRAS.373..906M}, using OASIS, an integral-field spectrograph with higher spatial resolution than SAURON, found a number of small KDCs, typically of only a few hundred parsecs in diameter. These KDCs were actually found in fast rotators and seem to be only counter-rotating (oriented at 180\degr\, with respect to the main body). The same study showed that these small KDCs should be considered different from the large KDCs found (mostly) in slow rotators, as they have younger stellar ages, often associated with gas and dust. Moreover, the authors argued that the small KDCs are essentially the product of recent accretion of relatively minor amounts of gas, which rejuvenates the nuclei, and the KDC is visible temporarily only because the mean velocity is a luminosity weighted measurement, and as such is biased by bright young stars. As the stellar population of the small KDC stars ages, its imprint on the observed kinematics on the host body will diminish and eventually disappear. 

The large KDCs are made of old stars, with colour or stellar population gradients similar to other early-type galaxies of the same mass \citep{1988ApJ...327L..55F,1997ApJ...491..545C,1997ApJ...481..710C,2001ApJ...548L..33D,2010MNRAS.408...97K,2015arXiv150103723M}, but their origin is not completely understood. The distinct nature of the angular momentum vector of the KDC with respect to the rest of the host galaxy suggests a KDC of external origin, and an evidence for mergers. The observation that the surface brightness profile of NGC\,5813 can be described with two components \citep{1980MNRAS.193..931E}, was generalised by \citet{1984ApJ...287..577K}, who suggested that the KDC in NGC\,5813 is a remnant of a small galaxy, and a similar argument could be applied to other KDCs. The light profiles of KDC host galaxies are, however, not different from other slow-rotators of similar masses \citep{1995AJ....109.1988F,1997ApJ...491..545C,1997ApJ...481..710C}, and the similarity of stellar populations within the KDC and the outer body of the galaxy is not consistent with this 'core-within-a-core' scenario \citep{1988ApJ...327L..55F}. In particular, if cores originated as small ellipticals, one should observe positive metallicity gradients (core would be metal poorer than the rest of the galaxy), but the opposite to this is observed in KDC hosts \citep[e.g.][]{1985MNRAS.215P..37E, 1988ApJ...327L..55F,1992A&A...258..250B, 1998A&A...332...33M, 2001ApJ...548L..33D, 2010MNRAS.408...97K}.
 
Other possible scenarios for the formation of KDCs include the disruption of a satellite galaxy during a minor merger, where the angular momentum of the KDC is the consequence of the orbital angular momentum of the satellite \citep{1988ApJ...327L..55F}, a dissipative merger of gas rich (spiral) galaxies, or accretion of external gas, where in both cases some of the acquired gas settles into the nucleus of the main progenitor and forms stars \citep{1992A&A...258..250B}. While KDCs seem to be indicators of past merger history, \citet{1988ApJ...330L..87J} pointed out that photometric evidence (i.e. shells) are not often associated with KDCs. This, however, is consistent with an early formation (supported by the old ages of stars within KDCs), where galaxies are built through hierarchical merging of small bodies, possibly partially gaseous, at an early epoch \citep{1992A&A...258..250B}.

The formation of KDCs has to be considered within the cosmological picture of galaxy formation, namely that there seems to be two phases of mass assembly for massive galaxies \citep[e.g.][]{2010ApJ...725.2312O}, where the first one is dominated by vigorous star formation fed by cold streams \citep{2005MNRAS.363....2K,2009ApJ...703..785D}, followed by the accretion of already formed stars from satellites \citep[e.g.][]{2007ApJ...658..710N,2011ApJ...736...88F,2012ApJ...754..115J, 2012MNRAS.425..641L}. 

There is one other group of kinematically complex galaxies, which are found on both sides of the slow -- fast rotator division line, but make up only 4 per cent of the total population of early-types in the nearby Universe. These galaxies were named by \citet{2011MNRAS.414.2923K} {\it double-sigma} galaxies, or $2\sigma$, because of unusual velocity dispersion maps. Velocity dispersion maps of $2\sigma$ galaxies are characterised by two symmetrically off-centered peaks along the major axis. The velocity maps can, however, have a variety of structures: from no net rotation to counter-rotating cores, and even a multiple reversal of rotation direction \citep[for examples of velocity and velocity dispersion maps of $2\sigma$ galaxies see figs~C.5 and C.6 of][]{2011MNRAS.414.2923K}.

The $2\sigma$ galaxies are simpler systems than KDC hosts, and their internal structure is better understood. They are also typically less massive \citep[$<5\times10^{10}$ M$_\odot$,][]{2013MNRAS.432.1709C} than galaxies with KDCs, such as NGC\,5813, but KDCs are also found in galaxies populating the same range in mass as $2\sigma$ galaxies. The two-dimensional spectroscopy of these galaxies revealed a double-peaked structure in the line-of-sight velocity distribution (LOSVD) indicating they are made of two separate kinematics components which counter-rotate \citep[e.g. NGC\,4550, ][]{1992ApJ...394L...9R, 1992ApJ...400L...5R}. Dynamical models based on  the orbit superposition have shown that such galaxies are indeed made of two counter-rotating components, both having high angular momenta \citep{2007MNRAS.379..418C}. Furthermore, the two components are characterised by different stellar populations \citep[e.g.][]{2011MNRAS.412L.113C,2013A&A...549A...3C,2013MNRAS.428.1296J}. Given the flatness of these galaxies, and the high angular momentum of each separate component, $2\sigma$ galaxies are consistent with being made of two counter-rotating discs of comparable masses, where the details of their kinematic features (rotation or no rotation in the velocity maps, the shape and the separation of the two peaks in the velocity dispersion maps) depend on the relative masses and characteristic sizes of the discs. For example, NGC\,4550 seems to be made of equally massive discs, while the counter-rotating discs in NGC\,4473 are approximately split 30-70 per cent in mass \citep[e.g.][]{2007MNRAS.379..418C}.  

The origin of $2\sigma$ galaxies is currently associated with accretion of gas, where the accreted external gas settles in the equatorial plane, but with an opposite angular momentum. The source of the gas can be a nearby object, such as in the case of NGC\,5719/NGC\,5713 pair \citep{2007A&A...463..883V, 2011MNRAS.412L.113C}, or, as recent numerical simulations suggest, a $2\sigma$ galaxy could sit in a locus of two cold streams, which feed the galaxy with gas and exert different torques resulting in the formation of the prograde and retrograde disks of  relative contributions proportional to the properties of the streams \citep{2014MNRAS.437.3596A}.

In this paper, we study NGC\,5813 with MUSE (Bacon et al. in prep.), focusing on its KDC, and the fact reinforced by new observations that its KDC is the result of the superposition of two counter-rotating structures, similar to what is found in $2\sigma$ galaxies. This has strong implications on the formation history of this kind of galaxies and the origin of (at least some) KDCs in massive slow rotators. We argue here that the KDC in NGC\,5813 is indeed an orbital composite, in a similar way as found in NGC\,4365 by \citet{2008MNRAS.385..647V}.

We start with a summary of general properties of NGC\,5813 (Section~\ref{s:gen}) relevant for this study. The observations and data reduction are presented in Section~\ref{s:obs}. The main evidence supporting our picture are the MUSE stellar kinematics presented in Section~\ref{s:skin} and the Schwarzschild dynamical models from Section~\ref{s:dyn}.  In Section~\ref{s:gkin} we present emission-line kinematics from MUSE obeservations, which show that the present gas distribution is not related directly to the KDC. We summarise the main results and discuss the possible origin of the KDC in NGC\,5813 in Section~\ref{s:discus}. Throughout this paper we assume NGC\,5813 is at a distance of 31.3 Mpc \citep{2001ApJ...546..681T}. 

%
%

\section{General properties of NGC\,5813}
\label{s:gen}

NGC\,5813 is one of the two brightest galaxies in the NGC\,5846 group \citep{2005AJ....130.1502M}. This is an isolated group which consists of three parts dominated by NGC\,5846, NGC\,5831 and NGC\,5813. Spectroscopic observations \citep{1982MNRAS.201..975E,1983ApJ...266...41D,1993MNRAS.265..553C,2004MNRAS.352..721E} revealed that NGC\,5846 shows very low mean motion, while NGC\,5831 and NGC\,5813 both harbour KDCs. NGC\,5846 and NGC\,5813 have similar K-band magnitude \citep[-25.01 and -25.09, respectively,][]{2011MNRAS.413..813C}, while \citet{2013MNRAS.432.1709C} derived masses of 3.7 and $3.9\times10^{11}$M$_\odot$, respectively (constrained within the SAURON observations). NGC\,5831 is a smaller and a less massive system. 

The optical imaging and spectroscopy show that NGC\,5813 is unusual both kinematically and photometrically. The light profile in the observations of \citet{1982MNRAS.201..975E} indicated a "two-component structure" and was markedly different from other elliptical galaxies. There are a number of ways one could parameterise such a  light distribution, and this is evident in the variety of approaches in the recent work. 

The nuclear region can be fit by a Nuker double power law \citep{1995AJ....110.2622L}, but also with a core-S\'ersic \citep{2003AJ....125.2951G}, as done by \citet{2011MNRAS.415.2158R}. Outer regions (beyond 2\farcs5) are well fit with a combination of a S\'ersic and an exponential \citep{2013MNRAS.432.1768K}, while fitting the HST imaging \citet{2012ApJ...755..163D} and \citet{2013AJ....146..160R} agree that the best fit is achieved through a combination of a S\'ersic and a core-S\'ersic profiles, where the core-S\'ersic has a low S\'ersic index (n=2.8 or 2, respectively), while the outer profile is close to exponential. Therefore, the outer profile of NGC\,5813 is remarkably similar to those found in disc galaxies. This outer exponential profile led \citet{2014MNRAS.444.2700D} to classify NGC\,5813 as an S0 galaxy. 

The isophotes show a change in both their flattening and their orientation \citep[e.g.][]{2014MNRAS.444.2700D}. Not considering the behaviour in the sub-arcsecond region, the change in ellipticity is prominent beyond 10\arcsec, roughly where the KDC stops, and changes from being approximately constant at 0.1 (within 10\arcsec) and continuously increasing to about 0.3 in the outskirts. The position angle exhibits a slightly more complex behaviour. Within the central 20\arcsec, it continuously changes with radius (within about 15\degr), but there is a noticeable difference between the KDC region ($<10$\arcsec) where it reaches a maximum (about 150\degr) and the outer region where it flattens to about 135\degr. 

\citet{1997ApJ...481..710C} and \citet{2001AJ....121.2928T} detected dust in the central region of NGC\,5813. The dust is observed in two distinct morphological configurations, as filaments and what appears to be a nuclear disc. The filaments are similarly distributed like the ionised gas converging on the nucleus, while the dust disc, a thin slab of dust about 1\arcsec\, in size, cuts the nucleus in half and it is oriented at $\sim148$\degr. This orientation is only a few degrees different from the orientation of the KDC \citep{2011MNRAS.414.2923K}, as well as the nuclear disc-like kinematic feature seen in the ionised gas, which we present in Section~\ref{ss:smallgas}. The V-I HST colour images indicate that what appears to be a double nucleus in NGC\,5813 is likely a consequence of dust obscuration \citep{1997ApJ...481..710C}. 

The sub-group of NGC\,5813 is rich in diffuse X-ray emission. \citet{2005AJ....130.1502M} find that X-ray emission profile of NGC\,5813 differs from the one of NGC\,5846, in the sense that it is less steep, more luminous and more extended. This could be interpreted as a consequence of different formation histories, where NGC\,5846 experienced more interaction with smaller galaxies \citep[for evidence of a recent merger see][]{1998A&A...330..123G}. 

The AGN activity of NGC\,5813 is currently not high. It is detected in NVSS \citep{1998AJ....115.1693C} at 1.41 GHz, as well as at 8 GHz \citep{2002A&A...390..423K}. At the spatial resolution of the latter study (2-3\arcsec) the radio source is unresolved. Perhaps the most telling evidence of past, but also quite recent AGN activity are the X-ray observations analysed by \citet{2011ApJ...726...86R}, which present a case for three buoyant cavities, lifted by the AGN jet which operated in time intervals of 90, 20 and 3 million years. The cavities rise approximately along the minor axis of the stellar body, where the emission-line gas is also located, as detected on the H$\alpha$ narrow-band observations \citep{1994A&AS..105..341G, 2011ApJ...726...86R,2014MNRAS.439.2291W} and in various ionised emission-lines \citep[Section~\ref{s:gkin},][]{2006MNRAS.366.1151S}, as well as in 100K cold gas \citep{2014MNRAS.439.2291W}.

%
%
\section{Observations and Data Reduction}
\label{s:obs}

MUSE\footnote{Some illustration of MUSE capabilities are found in \citep{2014Msngr.157...13B}.} is a general purpose second generation instrument operating at UT4 of ESO's Very Large Telescope. It is an integral field spectrograph with a field-of-view of 60\arcsec$ \times$ 60\arcsec, covered by 24 individual spectrographs, each supplied with a slicer of 48 mirrors arranged in $4\times12$ stacks. This configuration allows a wide spectral range (we use the nominal 4800 - 9300 \AA), a spatial sampling of 0\farcs2, and a spectral resolution ranging from $R = 1500 - 3500$ across the wavelength domain. It clearly supersedes other optical spectrographs that were used to observe our target.

NGC\,5813 was observed during the first commissioning run on February 11, 2014, in fair weather conditions, with recorded DIMM seeing ranging between 0\farcs6 and 0\farcs8 during the total 1.4h of integration. An {\it a posteriori} estimate of seeing using the 'white light' reconstructed image on a few globular clusters found in the field suggests that the final point-spread-function (PSF) had a full-width-at-half-maximum (FWHM)$<0\farcs7$. The exposure sequence comprised four on-source 20 min integrations and two offsets to observe a blank sky region, each of 2 min in duration, positioned about 7\arcmin\, in the south-west direction (along the minor axis of the galaxy). The observing sequence was: S-O-O-O-S-O, where O is an object (on-target) and S is a sky (off-target) observation. Each on-target observation was dithered, but not rotated. The data were reduced using the v1.0 version of the pipeline. 

The data reduction follows the standard steps adapted for MUSE and described in Weilbacher et al. (in prep)\footnote{A recent and short description is given in \citet{2014ASPC..485..451W}}. The pipeline works on pixel tables, large tables with information on all pixel values, corrected for bias and flat-field and their location on the CCD. Master bias, flat field and arc calibrations solutions are created using a set of standard calibrations observed on the same night. Each exposure (on- and off-target) was reduced separately, but in the same way, consisting of a subtraction of bias and flat-fielding. After determining the trace of the spectra, geometric and wavelength calibrations are used to transform the location of the spectra on the CCD into the focal plane spatial coordinates (in arcsec) and wavelengths (in \AA ngstr\"om). This was followed by the application of an astrometric solution. The standard star GD\,108 was observed at the beginning of the same night and prepared within the MUSE pipeline. Notably, the response curve was smoothed using a piecewise cubic polynomial function to remove high-frequency fluctuations. 

Sky was removed for each on-target cube separately, using one of the two closest in time sky cubes. The method implemented in the MUSE pipeline consists of two steps. Firstly, the reduction software determines a representative sky spectrum (continuum and sky lines) from each of the sky exposures. These sky spectra are then adapted using the information on the line-spread function for each of the 24 MUSE channels and removed from the on-target observations. In particular, the first sky was applied to the first on-target observations, the last sky on the last two on-target observations, while the middle on-target observation was sky-subtracted using an average sky-continuum from the two sky observations. 

The last step in the reduction was the merging of individual observations. As there were only simple shifts between the observations, the shifts were reversed and pixel tables were combined, producing a science-ready data cube. Throughout the reduction the pipeline calculates and propagates the formal noise at each step.

%
%

\section{Stellar kinematics}
\label{s:skin}

\subsection{Binning of the MUSE datacube}
\label{ss:bin}

Prior to extraction of the stellar kinematics, the MUSE cube of NGC\,5813 was spatially binned using the Voronoi tessellation method of \citet{2003MNRAS.342..345C}. The target signal-to-noise ratio (S/N) was 130, where this ratio was estimated as the ratio of the flux and the square root of the pipeline produced variance at 5500 \AA. We also created a binned cube with a target S/N=70 for testing and for the extraction of emission-line kinematics (see Section~\ref{s:gkin}). 

For the reliable extraction of stellar kinematics it is often sufficient to have a lower S/N ($\approx 40$ per \AA), but we chose a relatively high S/N in order to improve on the systematics possibly introduced by the specific dithering pattern used during the observations. The median S/N per pixel at 5500\AA\, of the MUSE data cube is 10, with the central regions reaching 100. The final choice of the target S/N was a compromise between the fact that we wanted to limit the sizes of the central bins, such that they are smaller or comparable to the seeing disc, but large enough to provide sufficient signal for a robust extraction of the kinematic features.

\begin{figure*}
\includegraphics[width=\textwidth]{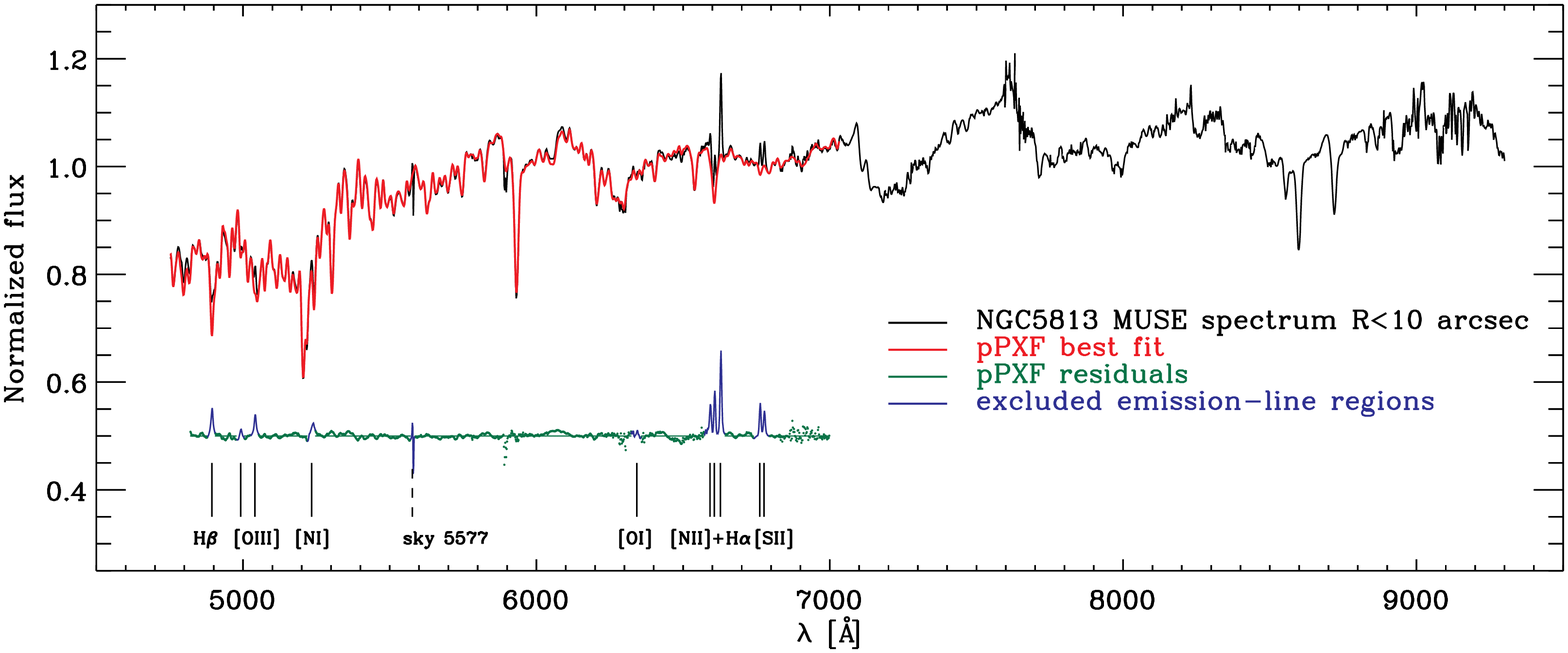}
\includegraphics[width=\textwidth]{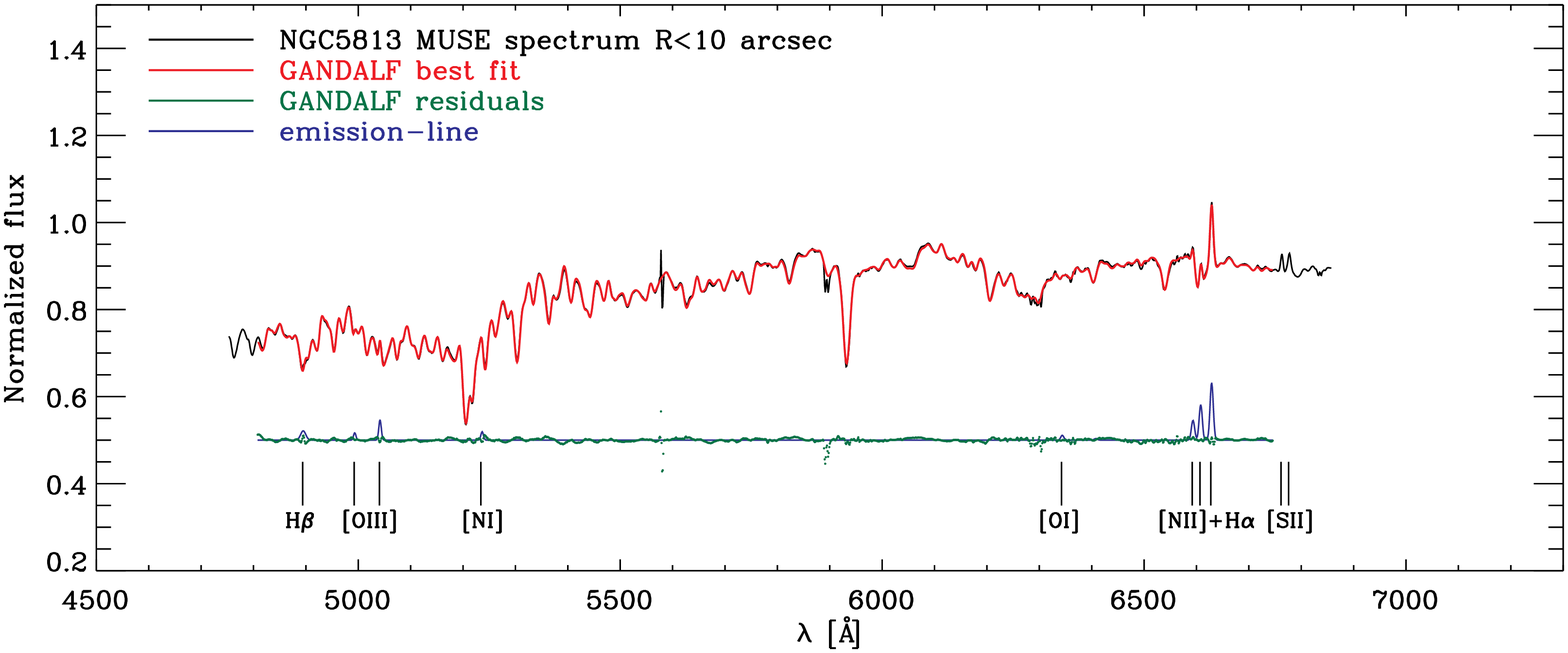}
\caption{MUSE spectra of the central region of NGC\,5813, integrated within a circular aperture of a radius of 10\arcsec, showing the pPXF (top) and GANDALF (bottom) fits. The bottom spectrum is shortened and shows only the fitted region. The fit is limited to the region blue ward of 7000\AA, as the spectra are significantly influenced by the tellurics and imperfect sky subtraction in the red. For the pPXF fit, we masked regions around the prominent emission (and sky) lines, but they are fitted with GANDALF.  Legend shows the components of the fit and the prominent lines of interest. The pPXF fit gives a S/rN of 200, while GANDALF fit provides a S/rN of 255. }
\label{f:ppxf}
\end{figure*}

\subsection{MUSE kinematics}
\label{ss:mkin}

We use the publicly available penalised Pixel Fitting (pPXF)\footnote{http://purl.org/cappellari/idl/} method of \citet{2004PASP..116..138C} to determine the LOSVD of stars. This method uses the Gauss-Hermite parametrisation of the LOSVD \citep{1993ApJ...407..525V,1993MNRAS.265..213G}, which are used to estimate departures in shape of the LOSVD from a Gaussian, such as the skewness and kurtosis. pPXF builds an optimal spectral template from a library of stellar (or galaxy) spectra, which when broadened with a given Gauss-Hermite function, will best fit the observed spectrum. We used the entire MILES\footnote{MILES library is available from http://miles.iac.es} library of stellar spectra \citep{2006MNRAS.371..703S,2011A&A...532A..95F}. The resolution of the stellar library (2.50\AA, or $\sigma=48$ km/s at 6560\AA) is smaller, but very similar to the resolution of the MUSE spectra, which show a changing resolution: at 5000\,\AA\, of 2.74 (69 km/s), at 6000\,\AA\, of 2.59 (54 km/s) and at 7000\,\AA\, of 2.54 (46 km/s). As the resolution is significantly lower than the stellar velocity dispersion in NGC\,5813, we did not convolve the spectral library to the same resolution of the MUSE data. 

\begin{figure*}
\includegraphics[width=\textwidth]{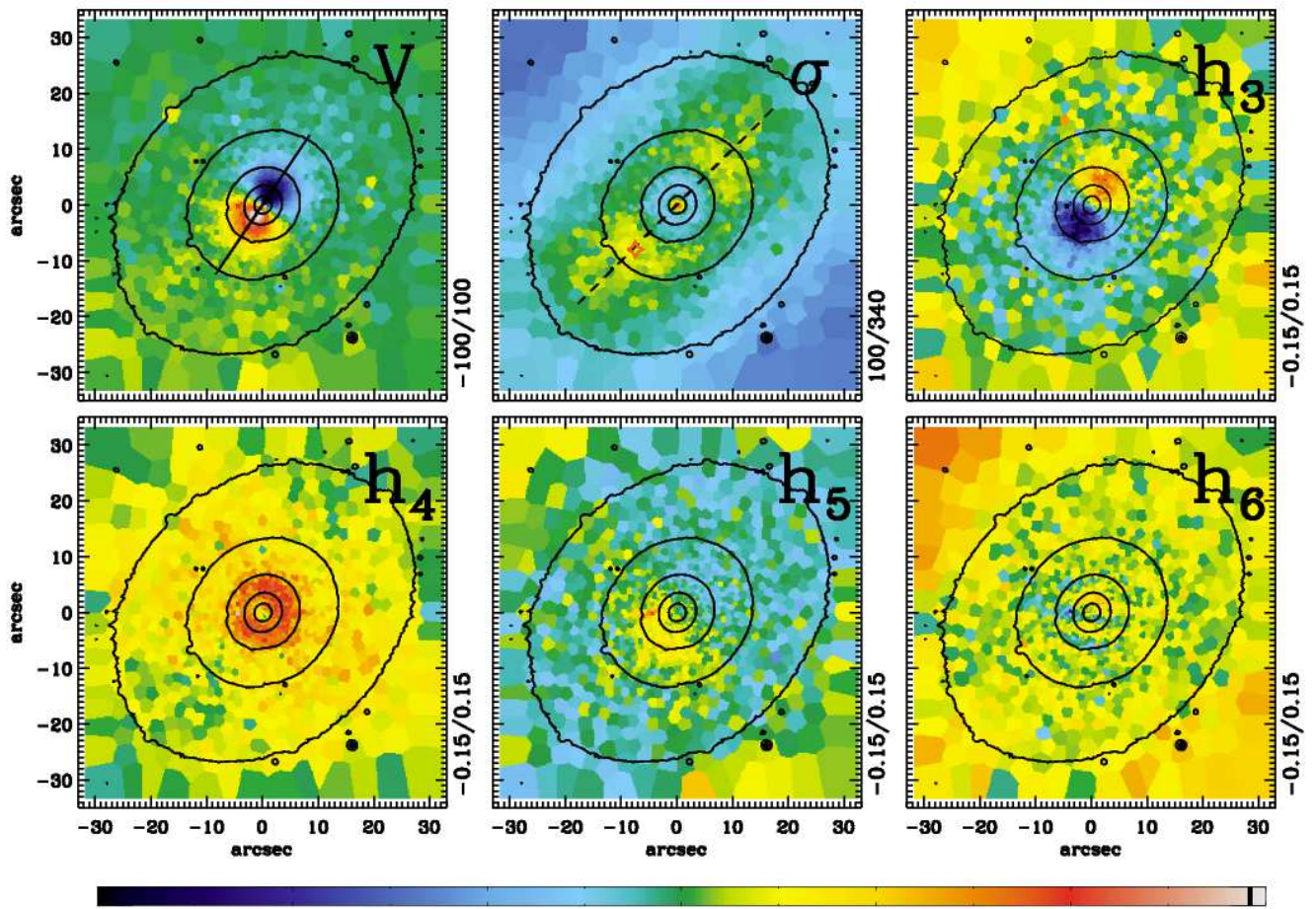}
\caption{Maps of LOSVD moments of NGC\,5813. Top row: the mean velocity, the velocity dispersion, Gauss-Hermite coefficient $h_3$. Bottom row: Gauss-Hermite coefficients $h_4$, $h_5$ and $h_6$. Colour scale is given by the colour bar at the bottom, while the ranges for each moment are shown on the lower-right corner of each map. When plotting the $h_3$ and $h_5$ maps we subtracted the medians of 0.012 and -0.015, respectively. The black contours are isophotes from the white light MUSE image.  The straight line on the mean velocity map shows the orientation of the KDC and the inner photometric major axis (PA=146\degr). The dashed line on the velocity dispersion map shows the outer (global) photometric major axis (PA=135\degr), which coincides with  the orientation of the $2\sigma$ peaks. The diamond symbol is the location of the bin used in Section~\ref{ss:2s} and Fig.~\ref{f:losvd}. North is up and East to the left.}
\label{f:Skin}
\end{figure*}

Both the library and the MUSE spectra were logarithmically rebinned and the pPXF fit was performed in two steps. The purpose of the first step was to derive the optimal template by fitting the full spectral library ($\sim1000$ spectra) to a representative spectrum, which was created by collapsing the full NGC\,5813 data cube. We set up pPXF using 6 moments of the LOSVD ($V$, $\sigma$, $h_3$, $h_4$, $h_5$, and $h_6$), and we added a 4th order additive polynomial. We used the default value for the penalisation (0.4). Finally, we limited the wavelength range used for the fit to $\lambda< 7000 $\AA, and excluded the regions with possible emission lines. We also excluded the region around the strong 5577\,\AA\, sky line which was not removed well for all spectra. A fit to a spectrum extracted from a circular aperture is shown on the top panel of Fig.~\ref{f:ppxf}. In order to verify the extracted kinematics, we fitted various spectral regions. For example, we fitted the full wavelength range and the SAURON wavelength range (4570 - 5200\,\AA) for comparison. These fits resulted in the same general trends and kinematic features. 

In the next step we run pPXF on the individual spectra of the MUSE data cube using always the optimal template constructed from the five spectra with non-zero weights assigned by the pPXF fit in the first step. For each MUSE spectrum we estimated S/rN, where the rN (residual noise) was estimated as the resistant standard deviation of the difference between the spectrum and the pPXF best fit, limited only to the wavelength region used in the fit (also excluding emission-lines). The central regions are characterised by S/rN of around 130, which then steadily drops to about 90 at the distance of 28\arcsec. 

The uncertainties were estimated running Monte Carlo simulations for each bin, where the original spectrum was randomly perturbed 500 times. The level of perturbation is determined from the rN value of the original pPXF fit to the spectrum. The only change to the pPXF set up was in the removal of the penalisation during the Monte Carlo simulations. Average uncertainties on $V$ and $\sigma$ are 3 km/s and 4 km/s, respectively, while for the $h_3$ to $h_6$ Gauss-Hermite moments they are $\sim0.013$. Errors are constant across the field with a dispersion of about 0.5 km/s for $V$ and $\sigma$, and 0.0025 for the Gauss-Hermite moments, with a small tendency for an increase in the outermost (edge of the field) bins. 

Figure~\ref{f:Skin} presents the kinematic maps of NGC\,5813: the mean velocity, velocity dispersion and the higher order Gauss-Hermite coefficients: $h_3 - h_6$. The MUSE data confirm what has been observed already before \citep{1982MNRAS.201..975E, 2004MNRAS.352..721E}: the KDC in the central 10\arcsec\, of the velocity map, followed by a zero level mean motion, and the complex structure in the velocity dispersion maps consisting of a central peak, a nearly circular region with a significantly lower $\sigma$ values around it, the subsequent rise and the final decrease of the velocity dispersion values. The $h_3$ map shows anti-correlation with the KDC velocity structures and, as expected, the $h_5$ map shows a correlation. The median values of $h_3$ and $h_5$ maps are not zero, but 0.012 and -0.015, respectively. This is possible due to a minor template mismatch, but within the errors, the maps are consistent with zero outside the KDC region. For presentation purposes we removed the median values when plotting. $h_4$ has a relative drop in the very centre (associated with the central rise in $\sigma$), surrounded by an elevated region corresponding to the KDC area, which then falls off towards the edges. The  $h_6$ map is essentially flat. 

\subsection{Kinematic evidence against two thin counter-rotating discs}
\label{ss:2s}

\begin{figure}
\includegraphics[width=\columnwidth]{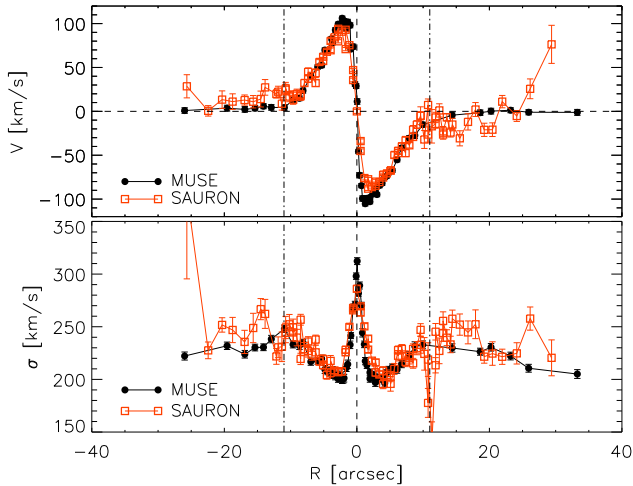}
\caption{Comparison of MUSE (black circles) and SAURON (red open squares) velocity and velocity dispersion values along the photometric major axis (at large scales).  Note the similarity of the two kinematics measurements, but the higher quality of MUSE data, both in terms of smoother variations and higher spatial resolution. }
\label{f:sau}
\end{figure}

In the case of NGC\,5813, the most striking is the structure on the velocity dispersion map. In Fig.~\ref{f:sau} we show the similarities between the SAURON and MUSE velocities and velocity dispersions along the major axis of NGC\,5813. The existence of the complex structure on the velocity dispersion maps is already clear on the SAURON kinematics \citep{2004MNRAS.352..721E}, but due to the shape of the field-of-view of the final SAURON mosaic, and the noise in the velocity dispersion map, it was not possible to recognise the specific shape of the high $\sigma$ structure. The MUSE data show that the region with the elevated velocity dispersion is symmetric and flattened along the photometric major axis (and not, for example, distributed in a circular ring). Furthermore, the velocity dispersion map consists of two distinct peaks along the photometric major axis of the galaxy (the velocity dispersion is lower along the minor axis). The peaks occur around 11\arcsec\, from the centre, coinciding with the edge of the KDC, as can be seen in Fig.~\ref{f:sau}.

This shape is in a striking resemblance to the shapes of the velocity dispersion maps found in $2\sigma$ galaxies, made of two counter-rotating discs \citep[for examples of velocity and velocity dispersion maps see figs~C.5 and C.6 of][]{2011MNRAS.414.2923K}. As discs counter-rotate, the combined LOSVD is broadened and deformed, where the relative speed of rotation might even induce a clearly separated two peak shape of the LOSVD \citep[e.g. as in NGC\,4550, see][]{1992ApJ...400L...5R}. In general, the broadening does not have to be sufficient for such a separation in two peaks (the relative speeds are not sufficiently different for such a separation), but instead manifests itself in a general broadening of the LOSVD along the major axis (e.g. as in NGC\,4473). The suppression of the mean rotation is also a direct consequence of the counter-rotation. If the two components have similar masses and distributions (e.g. disc thickness and scale lengths) then there will be no net rotation. If, however, one component is dominant within a certain region, the result will be an interchange of rotation, which can result in a counter-rotating core \citep[e.g. NGC\,0448][]{2011MNRAS.414.2923K}, or a disappearance of rotation in the central region (NGC\,4550) or at larger radii \citep[NCG\,4473,][]{2004MNRAS.352..721E,2013MNRAS.435.3587F}

\begin{figure}
\includegraphics[width=\columnwidth]{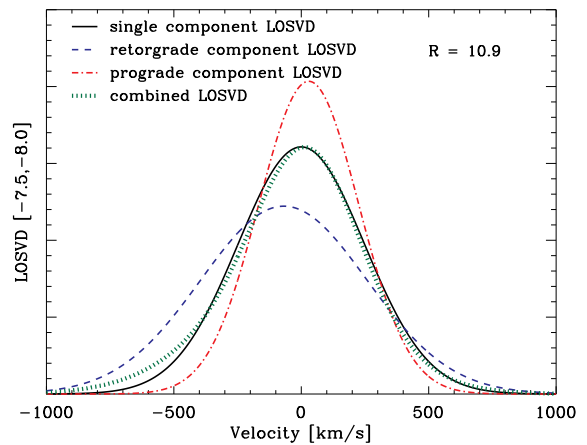}
\caption{Gaussian LOSVDs extracted at position (-7.5,-8.0) and 10.9\arcsec\, away from the nucleus in the south-east region of the high velocity dispersion are. All LOSVDs are normalised such that their integral is equal to one.  The single component LOSVD is shown by black full line, while the two separate LOSVD, pertaining to the prograde and retrograde motion are shown with blue (dashed) and red (dot-dashed) lines, respectively. A combined LOSVD (adding prograde and retrograde components) is shown by a green (dotted) line. The systemic velocity from a single component fit was removed from all LOSVDs. }
\label{f:losvd}
\end{figure}

NGC\,5813 is, however, probably not made of two counter-rotating, relatively thin, discs like a typical $2\sigma$ galaxy. A typical $2\sigma$ galaxy has ellipticity larger than 0.5, although there are cases with 0.3-0.4 ellipticites in the outer regions \citep{2011MNRAS.414.2923K}. NGC\,5813 shows a significant change in ellipticity with radius, being 0.1 within the KDC region and somewhat flatter at larger scales (ellipticity of 0.27). Furthermore, as seen on Fig.~\ref{f:Skin}, the KDC is oriented at approximately 145\degr, but the two peaks in the velocity dispersion are oriented at approximately 135\degr. These two orientations are also reflected in the varying position angle of the major axis, which changes from about 145\degr in the central 5\arcsec\, to 135\degr\, beyond the central 10\arcsec\, \citep[e.g.][]{2014MNRAS.444.2700D}. It is certainly telling that the orientations of the kinematic features are linked to those of the photometric body, but they are not compatible with two thin discs. Even if discs are misaligned (one at 135\degr\, and the other at 145\degr), it would be difficult to account for such significant cancellation of the observed mean motion outside of the KDC. If there are two counter-rotating components, they are likely not fast rotating discs. 

If the LOSVDs of NGC\,5813 are indeed made of two counter rotating components, it could be possible to extract each component separately, as for NGC\,4550 or some other galaxies \citep{2011MNRAS.412L.113C, 2013A&A...549A...3C}. In the cases where the disentangling of the components is successful, there is typically a substantial difference in stellar populations, and, even if co-rotating, a substantial difference in kinematics \citep[e.g.][]{2013MNRAS.428.1296J,2014MNRAS.441.2212F}. In this case, as is shown below, there seems to be no age difference between the prograde and retrograde kinematics components, while there is a certain difference in the metallicity. This makes the disentangling of the components more difficult, and therefore we attempted to separate the components only at one representative spatial location.

We used the pPXF ability for two component fitting and, to probe the stellar population differences, we made use of the single stellar population spectra from the MIUSCAT model library \citep{2012MNRAS.424..157V}, designated as safe\footnote{MIUSCAT library is available from http://miles.iac.es.}, spanning a range of ages 0.06 - 14 Gyr, and metallicities [Z/H]=-0.4 -- 0.22. We first ran pPXF in a single component mode, using the full MIUSCAT library (200 spectra), and determined which stellar templates were used in the fit. Our selected spectrum can be fitted with four MIUSCAT templates, which are all old (11-14 Gyr) and have a typical metallicity of either -0.4 or 0.2. From these we selected two with significant weights, and the largest difference in the metallicity (ages did not differ), assigned then to each component and ran pPXF in the two component mode. In all cases we used Gaussian LOSVDs. We selected a bin in the south-east region of the high velocity dispersion values (marked by a diamond in Fig.~\ref{f:Skin}). 

Resulting LOSVDs are shown in Fig.~\ref{f:losvd}. There are: the LOSVD of the single component fit (similar to the kinematics presented in Fig.~\ref{f:Skin}, but limited to only the first two moments), retrograde and prograde LOSVDs from the two component pPXF fits\footnote{We define the retrograde component as the one that has a systemic velocity lower than the single component fit.}, and a combined LOSVD of the retrograde and prograde components. As one would expect, the single component LOSVD is rather similar to a combined LOSVD of the retrograde and prograde components, and this explains the characteristics of the observed V and $\sigma$ maps. The two components are different in their velocity dispersions and mean velocities. The difference in the latter, for this spatial bin, is only about 100 km/s, and much less than the velocity dispersions (326 km/s and 196 km/s for prograde and retrograde components, respectively). The large velocity dispersions of both components and their relative low separation in the velocity indicate that NGC\,5813 is not made of two discs. We postpone the discussion on the implication of this result to Section~\ref{s:discus}, but stress that the process of spectral disentangling is highly degenerate, especially if there is no significant difference in stellar population, or the separation in velocities is small. Therefore, the results presented in Fig.~\ref{f:losvd} should be interpreted with caution.

%
%

\section{An approximate dynamical model of NGC\,5813}
\label{s:dyn}

If NGC\,5813 kinematics are a result of two counter-rotating structures, orbit-based dynamical models can be used to deduce the relative mass fractions of the components and the distribution of the orbits that constitute them. We use the \citet{1979ApJ...232..236S,1982ApJ...263..599S} orbital superposition method for constructing asymmetric galaxies, with the code defined in \citet{2006MNRAS.366.1126C} and \citet{2007MNRAS.379..418C}. Those papers explain the workings  of the code in detail, which we do not repeat here, but give a brief overview of the necessary steps and choices for starting parameters.

\subsection{Mass models and the Schwarzschild three integral method}
\label{ss:mge}

The essence of the Schwarzschild method is the construction of a library of orbits, possible in the gravitational potential defined by a mass model. The orbits from the library are then combined building a model galaxy, which is projected on the sky, calculating the observables comprising the mass distribution and the histograms of the LOSVDs at each position on the sky. These are compared to the original mass model and MUSE kinematic observations. The model is constrained using the non-negative least-square method \citep{1974slsp.book.....L}. Evidently, there are three crucial ingredients needed to robustly constrain the dynamical models: a mass model, a representative orbit library, and high-quality kinematics. 

We use the parametrisation of light of NGC\,5813 from Table B2 of \citet{2006MNRAS.366.1126C}, which is based on the Multi-Gaussian Expansion method \citep[MGE, ][]{1994A&A...285..723E} and constructed from an HST/WFPC2/F814W image and a ground-based image from the MDM observatory in the same filter. The step from  the MGE to a mass model is accomplished by deprojecting the observed surface brightness  using the formulas given in \citet{1992A&A...253..366M}, \citet{1994A&A...285..739E}, \citet{2002MNRAS.333..400C}, as well as the robust approximations from \citet{2002ApJ...578..787C}, assuming axisymmetry and a certain (constant) mass-to-light ratio, $M/L$. The stellar orbits are then calculated within the potential defined by the stellar density and a central massive black hole. For the deprojection we assume the inclination of 89\degr, but we also run a model with inclination of 60\degr, which gave essentially the same results. 

The orbit library needs to be sufficiently large to provide a complete set of basis function that can describe the distribution function of a galaxy. The fundamental problem is that the orbits are essentially $\delta$-functions while the distribution function is smooth.  \citet{2006MNRAS.366.1126C} resolved this issue by adding up a large number of 'dithered' orbits started from adjacent initial conditions.  Each orbit is defined with three integrals of motion: energy $E$, the $z$-component of the angular momentum $L_z$ and the third integral $I_3$. The orbits are here sampled on a three dimensional grid with 41 energy points (corresponding to the representative radius of the orbit), and at every energy we used 10 angular and 10 radial sectors, arranged in a polar grid, with the number of 'dithers' equal to six for each dimension of the grid \citep[for detailed visualisation of the orbital starting scheme and the dithering pattern see fig. 6 of][]{2006MNRAS.366.1126C}. Accounting for prograde and retrograde orbits, the complete orbit library consist of more than 1.7 million orbits. 

The potential within which the orbits are integrated is defined by the choice of symmetry, mass model and additional (dark) sources of gravity (central black hole and the dark halo). We run the axisymmetric version of the Schwarzschild code limiting the possible orbits to various sorts of short axis tubes \citep{1985MNRAS.216..273D}, but other resonant and chaotic orbits can occur. The observations presented in Sections~\ref{s:gen} and~\ref{ss:2s} indicate that the galaxy is at least weakly triaxial. Our approximation is, however, sufficient for the purposes of the present study, as we are not interested in the detailed shape of the galaxy. We want to test if the MUSE kinematics (the KDC and the $2\sigma$ peaks structure on the velocity and velocity dispersion maps, respectively) can be reproduced by two counter-rotating orbital families, which, given the orientation of the KDC, have to rotate around the minor (short) axis. As much as they could be present, the long-axis tubes and box orbits are not expected to contribute much to the observed kinematics, and we are not interested in determining the mass of the black hole (M$_{bh}$), dark matter contribution to the overall M/L, or the three dimensional shape of the galaxy. 

Finally, crucial constraints for a Schwarzschild model are provided by the MUSE kinematics (Section~\ref{s:skin}). As the model is by construction axisymmetric, we symmetries the MUSE data using point-(anti)symmetry around the global photometric axis (PA=135\degr), averaging the values at positions: [(x,y), (x,-y), (-x,y), (-x,-y)], but keeping the original errors for each point.

\begin{figure*}
\includegraphics[width=\textwidth]{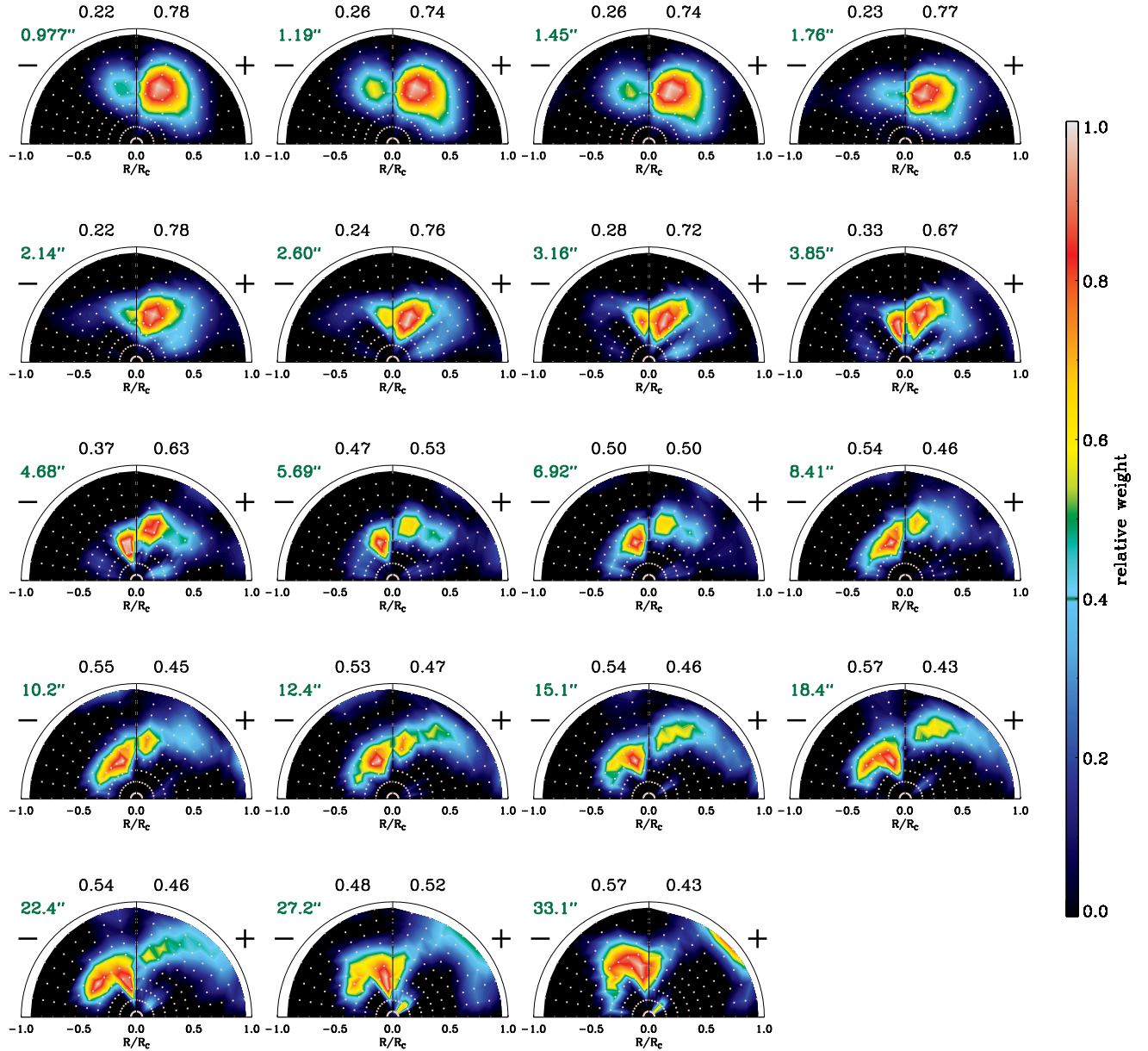}
\caption{Integral space of an axisymmetric orbit-superposition model for NGC\,5813. Each panel plots the meridional plane ({\it R,z}) for a given energy with the starting positions of orbits shown with white dots.  Energies are presented with the radius of the corresponding circular orbit in arcsec (green numbers in the upper left corner). Prograde orbits are shown on the right side (with a `$+$' sign), while the retrograde on the left side (with a `$-$' sign) of each panel. The radius of the circular orbit is the size of the horizontal axis (measured from the centre). Orbits with the highest angular momentum are found in the right and left corners. Over-plotted is the fraction of mass assigned to orbits at constant energy. The percentage of the total mass assigned to prograde and retrograde orbits are printed above the corresponding side of the panel. We show only a subset of radii, covering the KDC and the outer body fully within the MUSE observations.  The colour bar represents the relative colour coding of the orbital mass weights, where 1 represents the largest mass weight assigned to an orbit at the given energy. }
\label{f:orbit}
\end{figure*}

\subsection{Orbital space of NGC\,5813}
\label{ss:orbits}

We construct a Schwarzschild model of NGC\,5813 assuming M/L=4.81 \citep[as in][]{2006MNRAS.366.1126C} and a M$_{bh}$ from the recent M$_{\rm BH}$ - $\sigma$ relation of \citet{2013ApJ...764..184M}. Using the effective velocity dispersion of 210 km/s \citep{2013MNRAS.432.1709C}, the expected M$_{bh} = 2\times10^8$ M$_\odot$. At the distance of 31.3 Mpc this black hole will have a radius of the sphere of influence of less than 0\farcs2, which is below our resolution elements (the seeing was about 0\farcs7 FWHM). Although Schwarzschild models constrained with high quality IFU data can probe scales about 3 times smaller than the seeing resolution \citep[e.g.][]{2009MNRAS.399.1839K}, we do not fit for the M$_{\rm BH}$. Therefore, we are not concerned about the exact shape of the PSF, which we parameterise with a single Gaussian ($\sigma_{\rm psf}=0.3$\arcsec). We are primarily interested in the region between about 1-30\arcsec, the region dominated by the KDC and the outer, non-rotating parts of the galaxy which show the $2\sigma$ structure. We also ignore the contribution of the dark matter halo. The half-light radius of NGC\,5813 is about 57\arcsec, and the MUSE field-of-view, as large as it is, still samples just above a half of the effective radius. The contribution of the dark matter within this region is expected to be moderate ($\sim$15 per cent), but \citet{2013MNRAS.432.1862C} find about 47 per cent of dark matter within the half radius. Their Jeans anisotropic model \citep{2008MNRAS.390...71C}, however, has a quality flag 0, indicating a problematic data or model \citep{2013MNRAS.432.1709C}. The comparison of the model, which reproduces well all major kinematics features, with the data is shown in Fig.~\ref{af:model} of Appendix~\ref{a:model}. 

In Fig.~\ref{f:orbit} we show the orbital space of the Schwarzschild model. There are 19 panels, each showing a polar grid of orbital starting position for a given energy (represented with the radius of a circular orbit at that energy). The radii were chosen to sample the KDC and the surrounding region. Each panel is divided in the right and left side, corresponding to prograde and retrograde orbits, respectively. The definition of what is prograde and retrograde is arbitrary, and we assign the component responsible for the KDC to be the prograde component. High angular momentum orbits are found in the lower right and left corners of the panels. Orbits with an increasing $I_3$ values are found higher above the x-axis. The colour contours show the relative mass weights assigned to orbits in the construction of the model. 

Two important results can be seen in this figure. Firstly, in every panel there is a significant fraction of mass assigned into both prograde and retrograde orbits, but this fraction is changing with radius. Within the region of the KDC, and especially the region where the observed rotation curve is rising ($<$5\arcsec), the orbital space is dominated by the prograde orbits ($\sim 70$ per cent). This ratio diminishes towards the visible end of the KDC, where about half of the orbits have one and the other half the opposite spin. In the region of the $2\sigma$ peaks ($\sim$10 -- 25\arcsec), and at the largest radii probed by the model ($\sim$30\arcsec), the retrograde component is slightly more dominant. 

Secondly, the orbital space of NGC\,5813 lacks high angular momentum orbits, located at the right and left corners of the semicircles. Instead, at each energy, orbits are defined with both moderate $L_z$ and $I_3$ integrals. There is, however, a difference between the two components, where the retrograde orbits are characterised with a typically smaller angular momentum compared to the prograde orbits. In all panels the retrograde orbits are confined to the region with $R/R_c<0.5$, while this is not the case for the prograde orbits. This implies that the prograde component is overall rotating faster. This is also likely what drives the slight imbalance in the mass fractions beyond the KDC, where the net rotation needs to be zero. 

We run several models with different starting parameters: the inclination (also at 60\degr), the orientation of the symmetry axis (also at PA=145\degr , the orientation of the inner photometric major axis and the KDC), the black hole (an order of magnitude smaller and higher mass) and the M/L ratio (with $\Delta$ M/L=1). We also constrained the models with the kinematics extracted at the target S/N of 70 (same as used for the ionised gas extractions). All these changes gave rather similar orbital configurations always characterised by two components, where the ratio of the prograde to retrograde orbits changes from the initial 70:30 in the centre to 50:50 towards the end of the KDC region, while in the outer regions the orbital families had a 45:55 ratio. 

This implies that the best fitting orbital space is made of two, relatively thick, short-axis tube orbit families, which permeate the system. The structures are counter-rotating, but they are not thin discs made of stars at high angular momentum orbits with low third integral. Instead they are co-spatial vertically thick components, approximately identical in mass in the outer parts, but with the prograde component dominating in the centre (within the KDC). In Section~\ref{s:discus} we will discuss further the implication of this orbital structure.

%
%
\section{Emission-line kinematics}
\label{s:gkin}

NGC\,5813 is known to have an extended and morphologically complex distribution of ionised gas \citep{2006MNRAS.366.1151S, 2011ApJ...726...86R}, and it has a complex filamentary dust distribution with an apparent nuclear dust disk \citep[e.g.][]{1997ApJ...481..710C,2001AJ....121.2928T}. \citet{2014MNRAS.439.2291W} detect also cold (100 K) [{\sc Cii}] gas, which is co-spatial with the H$\alpha$ emission-line as well as the lowest entropy X-ray hot gas. The complex and multi-phase gas content of NGC\,5813 is most likely related to the cooling associated with the AGN activity cycles  \citep{2011ApJ...726...86R}. We investigate here if there is a connection between the gas content and the stellar KDC. A more thorough investigation of the emission-line gas properties we postpone to a future study.

\subsection{Distribution and kinematics of emission-line gas}
\label{ss:mgas}

\begin{figure}
\includegraphics[width=\columnwidth]{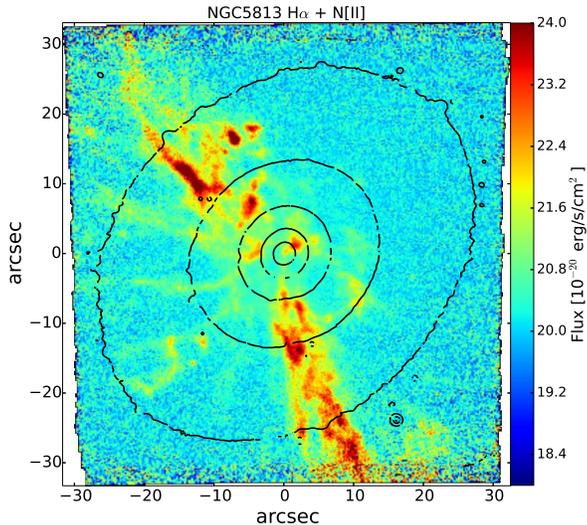}
\caption{MUSE 'narrow' band image of H$\alpha$  emission in NGC\,5813. The image as obtained by summing the flux within a narrow range around H$\alpha + \rm [{\sc Nii}]$ complex (between 6615 -- 6640 \AA), estimating the continuum level around that region and dividing it out. North is up and East to the left. Contours are the isophotes from the MUSE white light image. }
\label{f:halpha}
\end{figure}

 In Fig.~\ref{f:halpha} we show the H$\alpha$ + [{\sc Nii}] image summing the MUSE data cube between 6615 and 6640 \AA. The strong H$\alpha$ + [{\sc Nii}] features consist of two major filaments, one in the North-East direction, and the other in the South-South-West direction. Some other, less pronounced, filaments and knots are also visible, notably in the East part of the map. Other emission-lines typically follow the features seen here in H$\alpha$ + [{\sc Nii}] emission, although some are less prominent. Having the spectral information, we can proceed further and measure the kinematics of ionised gas. 

We use {\tt Gandalf} \citep{2006MNRAS.366.1151S, 2006MNRAS.369..529F} to measure fluxes and kinematics of the following emission-lines: H$\beta$, [{\sc Oiii}]$\lambda\lambda$4959,5007, [{\sc Ni}]$\lambda\lambda$5198, [{\sc Ni}]$\lambda\lambda$5200, [{\sc Oi}]$\lambda\lambda$6300,6364, H$\alpha$, [{\sc Nii}]$\lambda\lambda$6547,6583, and [{\sc Sii}]$\lambda\lambda$6716,6731. Fig.~\ref{f:halpha} shows that emission is abundant in certain regions and therefore we decided to change the binning from the high S/N used for the stellar kinematics. We again used the Voronoi-binning code of \citet{2003MNRAS.342..345C} setting the target S/N to 70 (again at 5500 \AA). This S/N is sufficient to robustly estimate the absorption features before attempting to fit the emission-lines, but it results in smaller bins: now we have more than 4000 spectra, which within the central 10\arcsec\, are rarely binned and have the original 0\farcs2 sampling. The central S/rN (of the pPXF fit) to these spectra approaches 140. 

\begin{figure*}
\includegraphics[width=\textwidth]{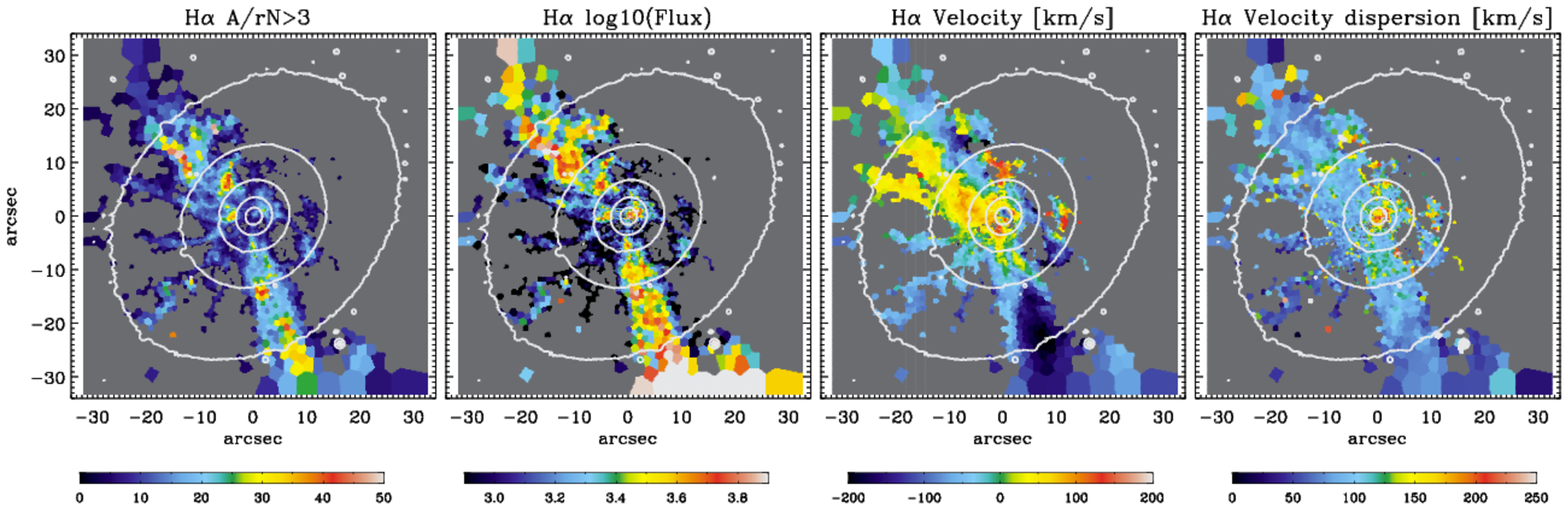}
\includegraphics[width=\textwidth]{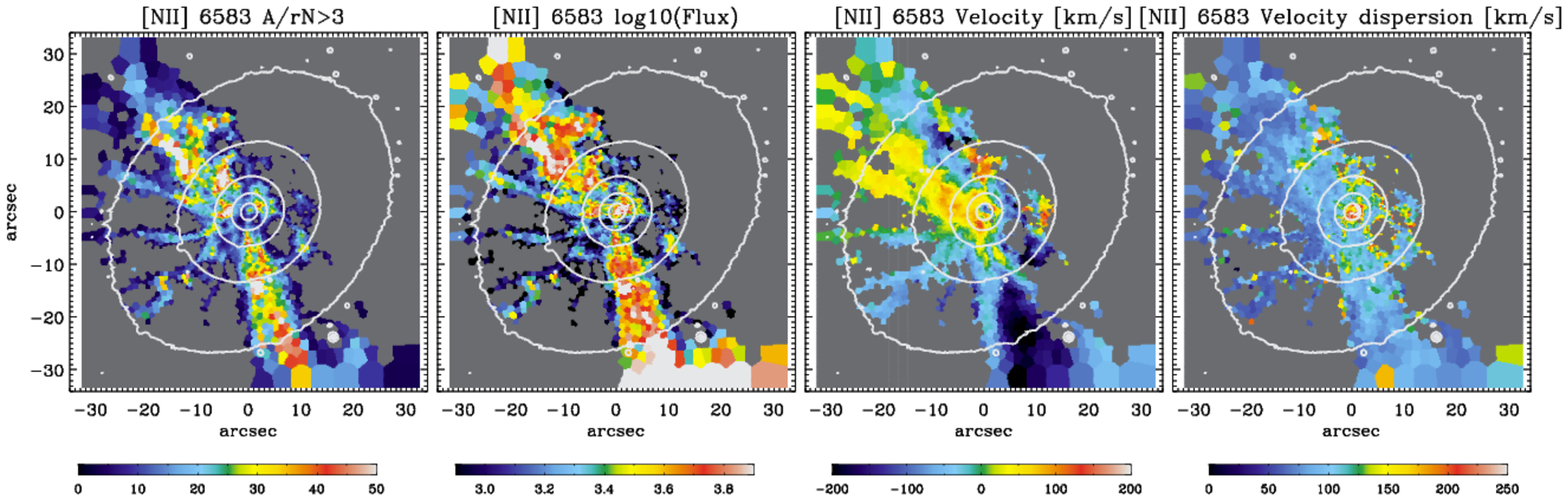}
\caption{Emission-lines H$\alpha$ (top) and [{\sc Nii}] (bottom) in NGC\,5813. From left to right: A/rN map where only emission-lines with A/N$>3$ are shown (region with no reliable emission is coloured grey), logarithm of emission-line flux (in units of 10$^{-20}$ erg/s/cm$^2$), the mean velocity and the velocity dispersion of a single Gaussian. Contours are the isophotes from the MUSE white light image. }
\label{f:gas}
\end{figure*}

The extraction of emission-line kinematics consisted first of a fit to the stellar continuum using pPXF and the MILES library, followed by a fit to the observed spectrum in which the optimal template (obtained from the first pPXF fit) is combined with several Gaussians mimicking the emission-lines. The call to pPXF differed somewhat from the one in Section~\ref{ss:mkin}, as we parameterised the LOSVD with only four moments and used the 10th degree multiplicative polynomial. The fitted region was the same as for the kinematics. An example of a GANDALF fit to a spectrum within a central circular aperture is shown on the lower panel of Fig.~\ref{f:ppxf}. 

We set up Gandalf to fit freely only H$\alpha$ and the stronger line of the [{\sc Nii}] doublet, while the kinematics of other forbidden lines were tied to the [{\sc Nii}] line and of H$\beta$ to H$\alpha$. This means that kinematic information could potentially be different only for H$\alpha$ and [{\sc Nii}] doublet, but fluxes of all lines were independently estimated. 

We show the distribution of all detected lines in the Appendix~\ref{b:emkin} (Fig.~\ref{bf:gas}), while in Fig.~\ref{f:gas} we show the fluxes, velocities and velocity dispersion for H$\alpha$  and [{\sc Nii}]. Only those bins for which the lines had an Amplitude-to-residual-Noise (A/rN) ratio\footnote{The amplitude refers to the amplitude of the fitted Gaussian.} greater than 3 are plotted \citep[following][]{2006MNRAS.366.1151S}.  With this criterion all of the fitted lines, except [{\sc Ni}], were detected, but [{\sc Oi}] was detected only in a very limited region, still following the general distribution of the [{\sc Nii}] line. The distribution of H$\alpha$ or [{\sc Nii}] on Fig.~\ref{f:gas} closely resembles what is seen on the reconstructed image in Fig.~\ref{f:halpha}. Similar structure was also seen in the SAURON maps of [{\sc Oiii}] and H$\beta$ emission-lines, but at $\sim5$ times lower spatial sampling\citep{2006MNRAS.366.1151S}.

H$\alpha$ and [{\sc Nii}] lines show similar and complex kinematic structures. Compared to the systemic velocity of the galaxy, the ionised gas in the southern arm is receding, while the northern arm shows both receding (east side of the arm) and approaching (west side of the arm) material. The measured velocity dispersion is relatively uniform over both arms and other filaments, typically between 60 km/s and 120 km/s (note that the instrumental resolution at the wavelength of H$\alpha$ is about 50 km/s and was not removed from the maps). The highest velocity dispersion values, in the excess of 220 km/s are found slightly off-centred towards the west. 

Emission-line gas is distributed almost in a polar configuration, extending close to the minor axis of the galaxy, and co-spatial with direction of the radio emission and the location of buoyant cavities in the X-ray emitting gas \citep{2011ApJ...726...86R}. The distribution of the emission-line gas in the southern region is marked by two regions with no detections. These cavities in the emission-line gas are centred on approximately (5\arcsec,-5\arcsec) and (15\arcsec, -15\arcsec), and they are co-spatial with the two innermost cavities in the X-ray gas. Given that the cold gas is most likely produced by thermally unstable cooling from the hot phase, \citet{2014MNRAS.439.2291W} suggest that the filamentary emission-line gas is linked to the buoyant cavities and the jet activity, where the relativistic plasma interact with the filaments and drags them out from the centre. 

\begin{figure}
\includegraphics[width=\columnwidth]{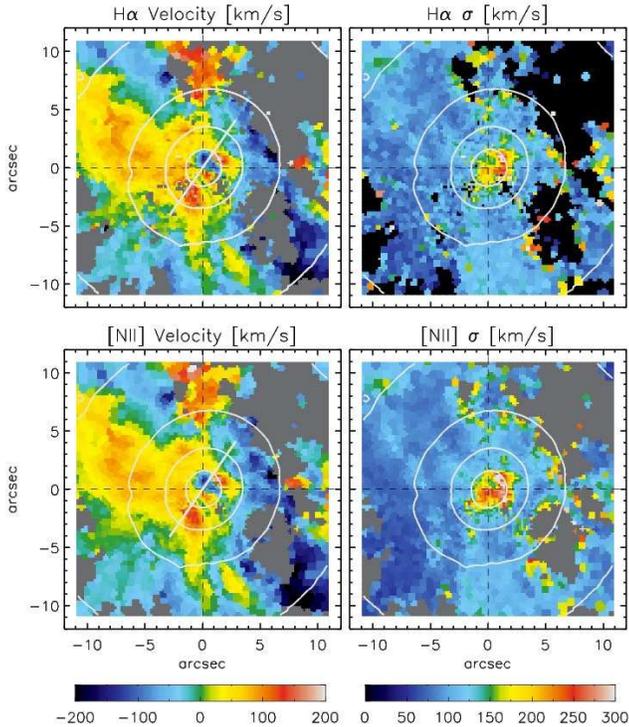}
\caption{Kinematics of H$\alpha$ (top) and [{\sc Nii}] (bottom) emission-lines within the central $10\times10$\arcsec. The mean velocity is on the left and the velocity dispersions on the right. The solid line on the velocity maps indicates the coinciding orientations of the stellar KDC and the dust disc visible on the HST images. The dashed vertical and horizontal lines indicate the position of the galaxy centre. Contours are the isophotes from the MUSE white light image. }
\label{f:gzoom}
\end{figure}

\subsection{Central emission-lines kinematics}
\label{ss:smallgas}

Both H$\alpha$ and [{\sc Nii}] emission-lines show peaks in the velocity dispersion in the very central region, not coinciding with the centre of stellar distribution (which is about 1\arcsec\, to the West). The peaks in emission-line velocity dispersion also coincide with a region which shows a different velocity structure of the ionised gas compared to the outer parts. Zooming in onto the central 10\arcsec$\times10$\arcsec, the different kinematics in the central 3\arcsec\, becomes apparent. We show this on Fig.~\ref{f:gzoom}. Coinciding with the nuclear dust disc seen on HST images \citep{2001AJ....121.2928T}, somewhat more extended ($\sim3$\arcsec), but with nearly the same orientation, there is a kinematic structure with approaching side in the North-West and the receding side in the South-East. The  approaching North-West side seems limited to the central 2\arcsec, surrounded by receding material. This interface between the approaching and receding material is also characterised by the peak in the velocity dispersion. The South-East side is more extended but also not uniform, showing two regions of increased velocities. This complex is present in both H$\alpha$ and [{\sc Nii}] emission-lines and displays similar distribution and kinematics. Given that it is spatially associated with the dust disc, it is possible that the velocity structure has the same inclination as the dust disc (nearly edge-on). The disc nature of the feature might, however, be difficult to reconcile with the off-centred high velocity dispersion peak and the non-regular spatial distribution of velocities mentioned above. It is likely that the observed motions are not solely of gravitational origin.

The higher S/N of MUSE data and the small size of the structure are possibly the reasons why this kinematic feature was not seen previously in the SAURON data \citep{2006MNRAS.366.1151S}. Alternatively, it could also be a feature that is associated more with H$\alpha$ and [{\sc Nii}] than with the lines observed within the short SAURON wavelength range (e.g. [{\sc Oiii}] which was used to derive the kinematics). In the MUSE data, H$\beta$ and [{\sc Oiii}] are essentially not detected in the very centre, although there is a hint of the kinematic structure on the [{\sc Oiii}] maps (see maps on Fig.~\ref{bf:gas}). This might also explain why \citet{2006MNRAS.373..906M}, who presented higher spatial resolution observations of SAURON sample galaxies with OASIS, also did not see this feature. 

Of particular interest is the fact that the central ionised gas rotation is oriented similar to the KDC and the nuclear dust disc (shown with a line on Fig.~\ref{f:gzoom}). We stress that the region of the high velocity dispersion in both H$\alpha$ and [{\sc Nii}] emission does not coincide with the centre of rotation, but surrounds the region of the approaching side of the central kinematic feature. As it arises where there is a reversal in the velocities, this could indicate shocks and the turbulent nature of the interstellar medium, linked with the relativistic plasma of the jets and the buoyant cavities.

\subsection{The multiphase gas, the activity and the KDC of NGC\,5813}
\label{ss:gas}

The central, possibly disc-like, kinematic structure revealed by MUSE is different from the rest of the ionised gas distributions. Its orientation is perpendicular to the radio jet and the direction of rising X-ray cavities, but it is oriented similar like the KDC. Is this gas feature related to the KDC? 

Stars within the KDC region are old and there is no evidence for any significant current star formation in this region \citep{2010MNRAS.408...97K}. The cold gas that is observed in NGC\,5813 is likely a result of cooling instabilities, which also seem to be typically present in cold-gas-rich systems \citep{2014MNRAS.439.2291W}. The spatial link between the orientation of the jet and the cavities filled with relativistic particles and the emission-lines gas detected in MUSE observations, suggests that these gas filaments are being dragged outwards by the jet and the bubbles. As described in \citet{2014MNRAS.439.2291W}, the hot particles can penetrate the filaments and heat the cold gas, both contributing to its distraction and creation, depending on their energy. 

If this scenario is correct, we are not seeing an inflow of the gas towards the centre, which settles in a disc-like structure and eventually creates stars and contributes to the stellar content and the kinematics of the KDC. On the contrary, we are seeing a gas reservoir which resides in the equatorial plane of the galaxy, and which could originate from the stellar mass loss, a relatively recent accretion or even be a remnant of a gas disc, which was, at an early epoch, responsible for the creation of the prograde rotating structure and the KDC. Currently this gas could be serving as the AGN fuel reservoir. The jet and the buoyant X-ray cavities are lifting and dragging the material from this gas reservoir, which is being cooled and forms a multiphase gas structure, as observed in the cold [{\sc Cii}] or warmer H$\alpha$ and [{\sc Nii}] emission. The velocity structure of the northern H$\alpha$ arm actually connects to the central gas disc, possibly supporting the evidence that gas is being lifted from the surface of the disc, but retaining its angular momentum. 

Therefore, the observations suggest that the present gas in NGC\,5813 is not of high relevance for the KDC. Further investigations of the properties of the emission-line gas, which could be used to asses this scenario, are beyond the scope of this paper, but we note that high spatial resolution observation of the cold gas phase centred on the nucleus would be beneficial. 

%
%

\section{Discussion}
\label{s:discus}

Previous sections presented MUSE observations of NGC\,5813 and their analysis. Here we summarise the evidence for the counter-rotating nature of the KDC and discuss the possible formation scenarios of NGC\,5813.

\subsection{The counter-rotating nature of the KDC in NCG5813}
\label{ss:evid}

The stellar kinematics of NGC\,5813 observed with MUSE suggests that this galaxy is made of two counter-rotating components. This is primarily based on the velocity dispersion map, which features an elongated structure with two peaks along the major axis of the galaxy, located some 11\arcsec\, from the nucleus. The appearance of the $\sigma$-peaks coincides with the visual end of the rotation and the KDC. In their spatial distributions, the velocity and the velocity dispersion maps are similar to maps of $2\sigma$ galaxies, for which it has been shown that they are made of two counter rotating structures. 

The stellar populations of NGC\,5813 are old everywhere \citep{2010MNRAS.408...97K}. The pPXF fit  based on stellar population models selects typically a few old stellar templates for reproducing the observed spectra, which, however, differ in metallicity. Fitting two templates of old ages, but different metallicities to a spectrum in the region of the $\sigma$ peaks, returns two velocity components, with mean velocities on the opposite sides of the systemic velocity for that spectrum. While this method is strongly degenerate, it suggests that there are two physical components with different sense of rotation. 

An orbit-superposition model, which fits the stellar kinematics well, provides further evidence for the two component nature of NGC\,5813. The two model components have opposite spins, where the prograde component typically has a higher angular momentum on average, Finally, the mass fractions assigned to the orbits vary with radius, but both orbital families are present everywhere. Within the KDC region, the prograde component is more dominant, explaining the KDC appearance, but in the outer regions the components contribute with essentially the same fraction of mass (small differences could be attributed to the difference in the angular momenta). 

The KDC region is also characterised by a drop in the velocity dispersion and a strong $h_3$ moment, which is anti-correlated with the mean velocity. This is typically seen in systems where the LOSVD can be constructed of two components, a fast and a slow rotating \citep{1994MNRAS.269..785B,1995A&A...293...20S}, both in cases of co- and counter-rotation. This is consistent with the orbital space of the Schwarzschild model, which shows that the prograde component within the KDC has a higher angular momentum than the retrograde component. 

While the evidence for two counter-rotating structures in NGC\,5813 are strong, it is also quite clear that these structures are not counter-rotating thin discs, like in a typical $2\sigma$ galaxy. NCG\,5813 is more massive than a typical $2\sigma$ galaxy. The shape of NGC\,5813 is also not as flat as in known $2\sigma$ galaxies. Moreover, the best fitting orbital space is not populated by high angular momentum orbits. Therefore, in spite of the fact that the outer light profile is best fit with an exponential (or almost exponential), the observed kinematics are not well described by a large-scale disc, or discs. No net rotation in the outer regions would be compatible with a disc only if it is seen exactly at an inclination of zero degrees (face-on). To show rotation, the KDC would then need to be made of a distinct component seen at a higher inclination. However, assuming the flattening within the KDC region ($\epsilon\sim0.1$) is due to an inclined disc, the disc would be viewed at an inclination of about 25\degr, while the outer disc would be required to be at about 45\degr\, ($\epsilon\sim0.3$). The stability of such a configuration even in a triaxial system is questionable. Furthermore, the two LOSVD components both have relatively high velocity dispersions, also arguing against the presence of thin discs.

A simple solution is that the KDC is an orbital composite, a result of a mixture of two components with different distribution of masses and similar, but not exactly the same, distribution of counter-rotating orbital families. A similar finding was already reported for another kinematically spectacular galaxy: NGC\,4365 \citep{2008MNRAS.385..647V}. This system is more complex than NGC\,5813, as the axes of rotation of the KDC and the outer body are perpendicular to each other, where the outer body is in a prolate-like rotation. Still, the triaxial models of van den Bosch et al. show that stars that make the KDC in NGC\,4365 are part of prograde short axis tubes which permeate the full system and the KDC is visible only due to an unbalance in the relative weights in the central 6\arcsec\, between the various orbital families. 

The origin of the unbalance in the known orbital distribution of KDC is not clear. A variety in possible formation processes for KDCs is highlighted by the range in the stellar populations \citep{1992A&A...258..250B, 1997ApJ...481..710C,2010MNRAS.408...97K} and kinematic properties \citep{2008MNRAS.390...93K,2011MNRAS.414.2923K} of galaxies with KDCs. What then are the possible formation scenarios for NGC\,5813 and its KDC?

\subsection{Assembly of NGC\,5813}
\label{ss:ass}

KDCs were heralded as show cases of merger driven formation scenarios. Galaxies harbouring KDCs do not differ from other, similar mass, slow rotator early-type galaxies. Still, there are significant characteristics of NGC\,5813 which point out to some possible formation pathways. 

The spatially uniform and old age of the stellar population in NGC\,5813 argues for an early assembly, or alternatively, a later assembly of the system comprising stars created at high redshift \citep[$z\sim3.5$,][]{2010MNRAS.408...97K}.  The negative metallicity gradient \citep{1985MNRAS.215P..37E, 1990MNRAS.245..217G, 2010MNRAS.408...97K}, constant abundance of the $\alpha$ elements relative to iron \citep{2010MNRAS.408...97K}, together with the finding in this paper that the two components permeate the (observed) body of the galaxy, argue against scenarios in which the KDC is a remnant of an accreted compact galaxy \citep{1984ApJ...287..577K,1988ApJ...327L..55F}. 

Furthermore, kinematic observations presented here argue against the origin of the KDC as a compact high-z elliptical which grows in size (and mass) via minor mergers, a scenario invoked to explain the rapid size evolution of quiescent galaxies \citep[e.g.][]{2009ApJ...699L.178N, 2010MNRAS.401.1099H}, but see also \citep[e.g.][]{2013ApJ...773..112C, 2013ApJ...777..125P, 2014ApJ...788L..29B} for an alternative scenario. A growth of NGC\,5813 via intermediate to minor mergers (e.g. mass ratios of 1:5 to 1:10) is, however, not supported by the outer exponential profile of this galaxy. \citet{2013MNRAS.429.2924H} show that such merger should increase the S\'ersic index to high values (n$>5$), comparable to those found in giant ellipticals. 

Processes that could be considered in the case of NGC\,5813 include:

\begin{enumerate}

	\item  Accretion of gas from a gas-rich companion which forms a disc-like structure in a principle plane of the system \citep[e.g.][]{1988ApJ...327L..55F}. This pathway is invoked for the creation of $2\sigma$ galaxies. An example is presented by \citet{2011MNRAS.412L.113C},  who show evidence for two counter-rotating discs in NGC\,5719, a galaxy which is stripping gas from its companion NGC\,5713 \citep{2007A&A...463..883V}. However, there is no strong kinematic evidence for thin discs in NGC\,5813, and NGC\, 5813 is a more massive system. Furthermore, the Schwarzschild model suggests a lack of high angular momentum orbits, and an existence of orbits with high vertical extent (above the equatorial plane). A viable pathway, however, is an early formation of the counter-rotating thin disc created from accreted gas, which due to subsequent interaction with satellites in the (sub)group environment was destroyed (dynamically heated and thickened). 

	\item  A major merger (mass ratios of 1:1) between two gas-rich spirals, or between a spiral and an early-type galaxy \citep[e.g.][]{1992A&A...258..250B}. Recent simulations \citep{2010ApJ...723..818H, 2011MNRAS.416.1654B, 2014MNRAS.444.1475M} show that binary mergers of galaxies, where at least one progenitor has a moderate gas content (e.g. 15-50 per cent), can produce KDCs and $2\sigma$ galaxies. Subsequent mergers of such galaxies can result in disruption of the KDC \citep{2011MNRAS.416.1654B}, or even an increased probability to host them \citep{2014MNRAS.444.1475M}. This should be taken into account as NGC\,5813, a central galaxy in a (sub)group, likely experienced a few similar mass mergers. The analysis of the orbital structure of the remnants by \citet{2010ApJ...723..818H} reveals that KDCs are indeed a consequence of the balance of relative contributions of orbital types spread throughout the system. \citet{2011MNRAS.416.1654B} show that if the spin of one of the progenitors is retrograde with respect to the orbital spin of the merger, the resulting KDC is also an orbital composite, with $2\sigma$ velocity dispersion features. Even initially prograde orientation of spins can result in a KDC, due to strong reactive forces which can cause the change in the orbital spin directly before the final coalescence \citep{2015arXiv150200634T}.  
	
	\item  A formation as a natural consequence of the the hierarchical clustering scenario, where the galaxy sits on an intersection of filaments bringing cold gas \citep{2005MNRAS.363....2K, 2009MNRAS.395..160K}. A scenario was put forward as a case for the creation of counter-rotating disc galaxies (essentially the 2$\sigma$ galaxies), where two streams are feeding the host dark matter halo with gas \citep{2014MNRAS.437.3596A}. If there is a sufficient difference in time between the arrival of the material (i.e. first through one filament and then through the other), two discs with opposite spins can be created. A simultaneous feeding would likely result in a single component due to the collisional nature of gas, if the gas has not already fragmented into molecular cloud complexes. This scenario is somewhat problematic for NGC\,5813 as there is no evidence for an age difference of stars. However, if the two feeding events happened early, were relatively short in duration and occurred within a few Gyr, it is essentially impossible to distinguish between the stellar populations. Moreover, given that NGC\,5813 is one of the two most massive galaxies in the NGC\,5846 group (and a centre of a sub-group) it is natural to imagine that the progenitor of the present day galaxy was formed on a location of merging cosmic filaments. The penetrating cold gas streams are rich in material and drive strong instabilities creating giant star forming clumps within the galaxy disc \citep{2009ApJ...703..785D}. Two streams, feeding the galaxy with material of opposite angular momenta could produce two thick and turbulent discs, which through a subsequent evolution could start resembling the two orbital families seen in NGC\,5813.		

\end{enumerate}

All these scenarios require gas to be present during the formation event. The gas-rich formation scenarios have difficulty accounting for a partially depleted core in NGC\,5813 nuclear surface brightness. The currently favourite scenario for creation of cores requires a binary black hole interaction in a dissipation-less merger \citep{1980Natur.287..307B, 2001ApJ...563...34M}. The core might have been created later, during the subsequent evolution of NGC\,5813 as a central galaxy in a group. The constraint to this scenario is that the merger (or mergers) can destroy the original disc-like structure, but needs to keep the counter-rotation present. Furthermore, the existence of fast rotators with partially depleted cores \citep{2013MNRAS.433.2812K} suggest there could be more unexplored pathways in creating kinematically disc-like structures and removing stars from the nuclei.

NGC\,5813 is a massive galaxy, containing stars consistent with being created early, within a short period of time. As a central galaxy in a (sub)-group it should have experienced both phases of the mass assembly \citep{2010ApJ...725.2312O}. In cosmological studies, galaxies with KDCs are still rare \citep{2007ApJ...658..710N, 2014MNRAS.444.3357N}, found in remnants of systems that went though a dissipation-less similar mass merger. When analysed, these systems have various types of orbits (short axis tubes, long axis tubes and box orbits), and the KDC is seen as an orbital composite \citep{2007MNRAS.376..997J,2014MNRAS.445.1065R}. Slow rotator merger remnants, however, do not fully resemble the observed galaxies: they are typically too flat and, in particular, the velocity maps of KDCs are not yet comparable to the observed systems \citep{2014MNRAS.444.3357N}. More work is required here, and a promising pathway in creating counter-rotating components (and KDCs in appearance) could be through cold streams which can create thick discs. The hope is that the halo mass and the properties of the streams will influence the outcomes, potentially creating a range of systems from less massive $2\sigma$ galaxies (with thin counter-rotating discs) to more massive, NGC\,5813-type galaxies with KDCs (with thick counter-rotating discs).

 The case of NGC\,5813 (together with NGC\,4365) suggests that KDCs could often be integral parts of the complete bodies of galaxies. It is the balance of orbits that makes KDCs visible. More KDC galaxies, spanning a range of masses, should be observed and analysed in a similar way to determine if NGC\,4365 and NGC\,5813 are typical or special objects, and if the KDCs are decoupled or integrated structures.  The spatial sampling, field-of-view, the spectral resolution and its efficiency make MUSE the right instrument to achieve this.

\section*{Acknowledgments}

DK thanks Norbert Werner for a useful discussion about the links between the observed emission-line gas and the large scale X-ray properties of NGC\,5813. DK thanks also Michele Cappellari and Richard McDermid for comments on an early version of the paper. Based on observations made with ESO telescopes at the La Silla Paranal Observatory under program ID 60.A-9100(A). R.B. acknowledges support from the ERC advanced grant 339659-MUSICOS.


\appendix

%
%

\section{Comparison of the Schwarzschild dynamical model and the MUSE kinematics}
\label{a:model}

\begin{figure*}
\includegraphics[width=\textwidth]{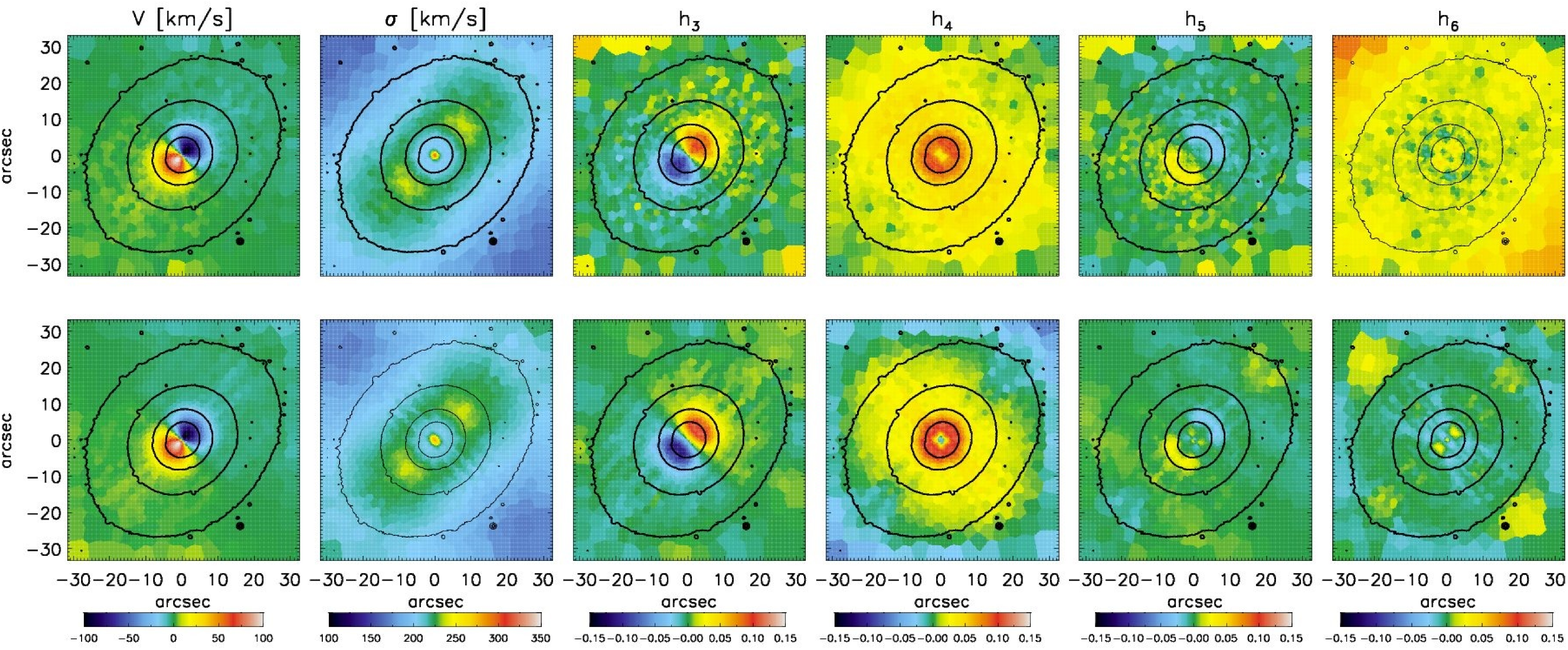}
\includegraphics[width=\textwidth]{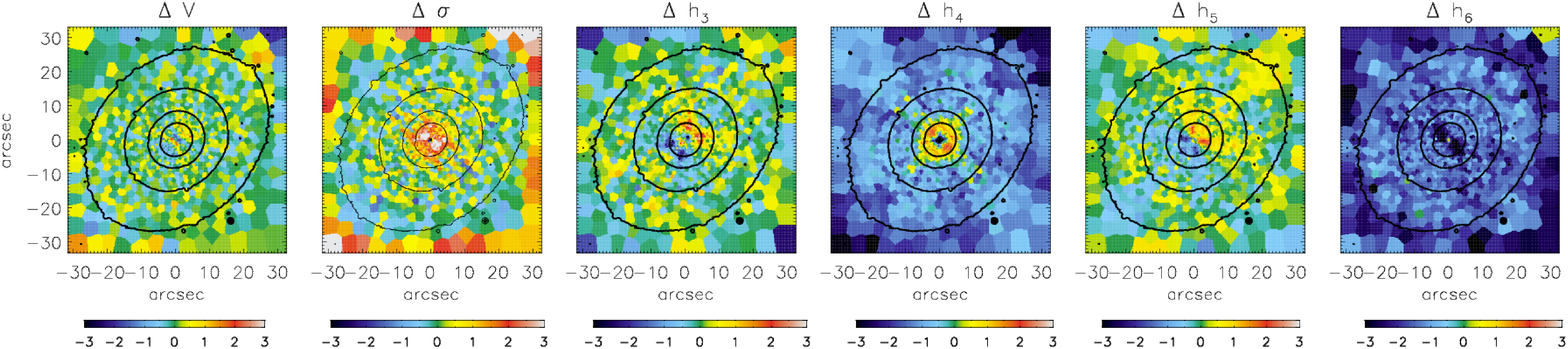}
\caption{Comparison between the MUSE kinematics (top) and the Schwarzschild model kinematic predictions (middle) from Section~\ref{s:dyn}. From left to right are shown: the mean velocity, the velocity dispersion and the higher-order Gauss-Hermite moments $h_3$ - $h_6$. The bottom row shows the residuals between the Schwarzschild model and the MUSE kinematics, divided by the observational errors.}
\label{af:model}
\end{figure*}

In Fig.~\ref{af:model} we show a comparison between the MUSE kinematics and the Schwarzschild dynamical model from Section~\ref{s:dyn}. The MUSE data were symmetrised using the point-(anti)symmetry as the model is by construction axisymmetric. The quality of the fit can be seen in the bottom row of maps which show the residuals between the model and the data, divided by the errors. The Schwarzschild model generally reproduced well the kinematics, with an exception of the centre, which suggests the model likely requires a higher black hole mass. 

\section{Other emission-lines detected within the MUSE wavelength range}
\label{b:emkin}

\begin{figure*}
\includegraphics[width=\textwidth]{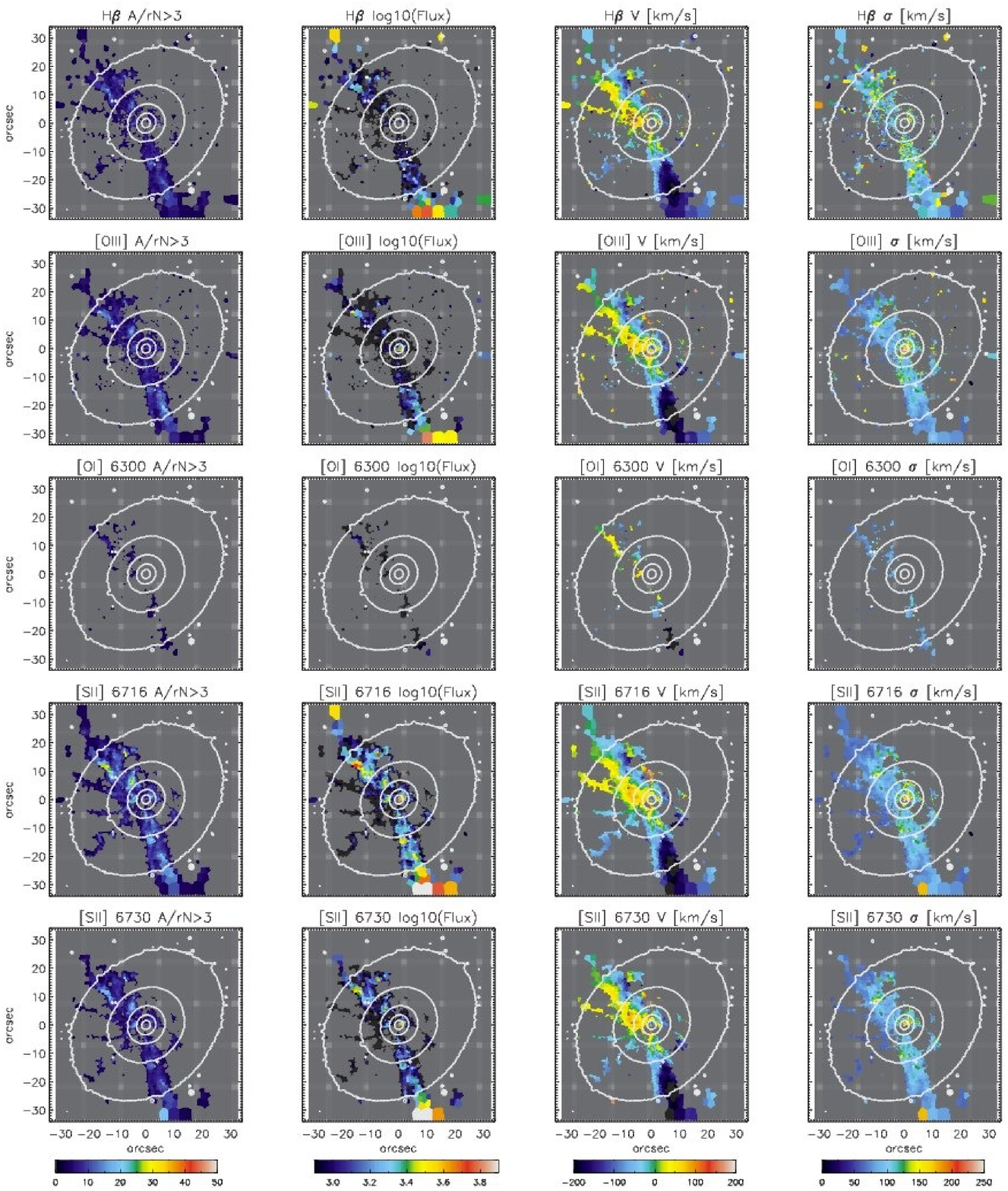}
\caption{From top to bottom: Emission-lines H$\beta$, [O\small{III}]$\lambda$4959,5007, [O\small{I}]$\lambda$6300, [S\small{II}]$\lambda$6716 and [S\small{II}]$\lambda$6731. From left to right: A/rN map, logarithm of emission-line flux (units of  of 10$^{-20}$ erg/s/cm$^2$), the mean velocity and the velocity dispersion of a single Gaussian. Extraction of these emission-lines is explained in Section~\ref{s:gkin}, but note that the velocity and velocity dispersion of H$\beta$ was fixed to that of the H$\alpha$ and of the forbidden lines to [N\small{II}]. This means that the lower four rows have the same kinematics, but different fluxes and distributions. Contours are the isophotes from the MUSE white light image. }
\label{bf:gas}
\end{figure*}

As described in Section~\ref{s:gkin}, we fitted a set of emission-lines typically found in the MUSE wavelength range:  H$\beta$, [{\sc Oiii}]$\lambda\lambda$4959,5007, [{\sc Ni}]$\lambda\lambda$5198, [{\sc Ni}]$\lambda\lambda$5200, [{\sc Oi}]$\lambda\lambda$6300,6364, H$\alpha$,  [{\sc Nii}]$\lambda\lambda$6547,6583 and [{\sc Sii}]$\lambda\lambda$6716,673. Of these only All lines expect  [{\sc Ni}] is not detected, and [{\sc Oii}] is detected on a much smaller area compared to the other lines. H$\alpha$ and [{\sc Nii}] are shown in the main text (Fig.~\ref{f:gas}, and in Fig.~\ref{bf:gas} we show the distribution and kinematics of all other deteceted lines.

\label{lastpage}

\end{document}